\documentclass{article}

\usepackage{arxiv}

\usepackage[utf8]{inputenc} 
\usepackage[T1]{fontenc}    
\usepackage{hyperref}       
\usepackage{url}            
\usepackage{booktabs}       
\usepackage{amsfonts}       
\usepackage{nicefrac}       
\usepackage{microtype}      
\usepackage{lipsum}         
\usepackage{amsmath}
\usepackage{cleveref}       
\usepackage{graphicx}
\usepackage{subfig}
\usepackage{multicol}
\usepackage[super,comma,sort&compress]{natbib}
\usepackage{doi}
\usepackage{xcolor}
\usepackage{lineno}
\usepackage[normalem]{ulem}
\usepackage{cancel}
\usepackage{float}
\DeclareMathOperator*{\argmax}{arg\,max}
\DeclareMathOperator*{\argmin}{arg\,min}

\usepackage{algorithm}
\usepackage{algpseudocode}

\title{Constraint-Guided Symbolic Regression for Data-Efficient Kinetic Model Discovery}

\author{\href{https://orcid.org/0000-0001-5273-7491}{\includegraphics[scale=0.06]{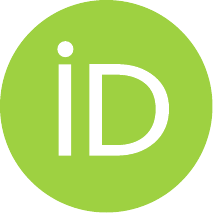}\hspace{1mm}Miguel Ángel de Carvalho Servia} \\
	Department of Chemical Engineering\\
	Imperial College London\\
    South Kensington, London, SW7 2AZ, UK\\
	\texttt{m.de-carvalho-servia21@imperial.ac.uk} \\
    \And
    \href{https://orcid.org/0000-0001-9648-5265}{\includegraphics[scale=0.06]{orcid.pdf}\hspace{1mm}Ilya Orson Sandoval} \\
	Department of Chemical Engineering\\
	Imperial College London\\
    South Kensington, London, SW7 2AZ, UK\\
	\texttt{o.sandoval-cardenas20@imperial.ac.uk} \\
 \And
  \href{https://orcid.org/0000-0002-1163-0505}{\includegraphics[scale=0.06]{orcid.pdf}\hspace{1mm}King Kuok (Mimi) Hii} \\
	Department of Chemistry\\
	Imperial College London\\
    White City, London, W12 0BZ, UK\\
	\texttt{mimi.hii@imperial.ac.uk}
  \And
 \href{https://orcid.org/0000-0002-4630-1015}{\includegraphics[scale=0.06]{orcid.pdf}\hspace{1mm}Klaus Hellgardt} \\
	Department of Chemical Engineering\\
	Imperial College London\\
    South Kensington, London, SW7 2AZ, UK\\
	\texttt{k.hellgardt@imperial.ac.uk}
 \And
   \href{https://orcid.org/0000-0001-5956-4618}{\includegraphics[scale=0.06]{orcid.pdf}\hspace{1mm}Dongda Zhang} \\
    Department of Chemical Engineering\\
    The University of Manchester\\
    Manchester, M13 9PL, UK \\
    \texttt{dongda.zhang@manchester.ac.uk} 
  \And
  \href{https://orcid.org/0000-0003-0274-2852}{\includegraphics[scale=0.06]{orcid.pdf}\hspace{1mm}Ehecatl Antonio del Rio Chanona $^*$} \\
	Department of Chemical Engineering\\
	Imperial College London\\
    South Kensington, London, SW7 2AZ, UK\\
	\texttt{a.del-rio-chanona@imperial.ac.uk}
}

\renewcommand{\headeright}{}
\renewcommand{\undertitle}{}

\begin{document}
\maketitle

\begin{abstract}
The industrialization of catalytic processes hinges on the availability of reliable kinetic models for design, optimization, and control. Traditional mechanistic models demand extensive domain expertise, while many data-driven approaches often lack interpretability and fail to enforce physical consistency. To overcome these limitations, we propose the Physics-Informed Automated Discovery of Kinetics (PI-ADoK) framework. By integrating physical constraints directly into a symbolic regression approach, PI-ADoK narrows the search space and substantially reduces the number of experiments required for model convergence. Additionally, the framework incorporates a robust uncertainty quantification strategy via the Metropolis-Hastings algorithm, which propagates parameter uncertainty to yield credible prediction intervals. Benchmarking our method against conventional approaches across several catalytic case studies demonstrates that PI-ADoK not only enhances model fidelity but also lowers the experimental burden, highlighting its potential for efficient and reliable kinetic model discovery in chemical reaction engineering.

\end{abstract}

\providecommand{\keyword}[1]{\textbf{Keywords:} #1}
\keyword{chemical reaction engineering, kinetic model generation, automated knowledge discovery, physics-informed machine learning, symbolic regression}

\pagebreak

\section{Introduction}\label{Introduction}
Catalytic processes are fundamental to industry, and their importance grows in the context of climate change and the urgent need to minimize waste while boosting efficiency. Kinetic models play a pivotal role in designing, optimizing, and controlling chemical reactors in these processes. The reliability of these systems fundamentally rely on the accuracy of the kinetic models, which capture the dynamic behavior of reactive system. Traditionally, model development has hinged on either mechanistic approaches, grounded in first principles and physical laws \cite{Baker_2018,Gernaey_2015}, or data‐driven methods, which leverage statistical and machine learning techniques. However, while mechanistic models are prized for their interpretability and theoretical grounding, they require significant domain expertise and are cumbersome to develop, but yet they are still widely established in industry and developed in research \cite{Jimenez_2011,Jedrzejewski_2014,Giessmann_2019}. Conversely, data‐driven models are flexible, easy to develop and can be faster to evaluate \cite{Sant_Anna_2017}, making them useful in real-time simulation \cite{Zhang_2019,DelRioChanona_2018, Park_2021, Sun_2022}, optimization \cite{Petsagkourakis_2020,RioChanona_2018,Wu_2023,Natarajan_2021}, and soft sensor development \cite{Mowbray_2022,Kay_2022,Kadlec_2009}. However, they often suffer from a lack of physical interpretability, may require large datasets to train (which are not always available in practice), and cannot easily extrapolate.

Symbolic regression is a method employed for automated knowledge discovery. Symbolic regression is a data-driven technique that seeks to identify interpretable and closed-form mathematical expressions which capture the underlying relationships in a particular dataset \cite{Haider_2023}. In recent years, symbolic regression techniques have become prominent tools for model identification, including ALAMO \cite{Wilson_2017}, SINDy \cite{Brunton_2016}, and genetic programming \cite{Koza_1994}. These methods can be broadly divided into two categories. The first category consists of evolutionary strategies, such as genetic programming, which only require variables and operators to be defined. This flexibility enables them to search an (almost) unconstrained space of candidate mathematical expressions without relying on predefined model structures. The second category comprises non-evolutionary approaches, exemplified by frameworks like SINDy (Sparse Identification of Nonlinear Dynamics) and ALAMO (Automated Learning of Algebraic Models for Optimization), which operate based on a design matrix that explicitly specifies the possible linear and non-linear transformations of the involved variables. In our earlier work on the automated discovery of kinetic rate models, we demonstrated the potential of genetic programming (in both its strong and weak formulations) to retrieve accurate kinetic models from sparse and noisy experimental data. Other notable works within this field are: Taylor et al.\cite{Taylor_2021}, Neumann et al. \cite{Neumann_2020}, Forster et al.\cite{Forster_2023}, Iba \cite{Iba_2008}, Nobile et al. \cite{Nobile_2013}, Datta et al. \cite{Datta_2019}, Sugimoto et al. \cite{Sugimoto_2005} and Cornforth et al. \cite{Cornforth_2012}.

Despite these advances, several challenges persist. One of the primary limitations is that the candidate models generated by a conventional genetic programming framework are sometimes physically implausible, lacking consistency with established chemical/physical principles (for example, ensuring that concentrations are always equal or greater than zero) \cite{deCarvalhoServia_2024, Forster_2024}. Additionally, these methods provide model candidates that give point estimates for model predictions, with no information regarding the uncertainty associated with those predictions: a factor that is critical in applications where safety and robustness are important.

This paper extends the work presented in our previous article by integrating two key enhancements into the automated discovery framework \cite{deCarvalhoServia_2024}. First, we incorporate mathematical constraints directly into the genetic programming algorithm. These constraints serve as a means to embed expert knowledge into the model generation process, effectively guiding the search towards solutions that are not only statistically optimal but also physically meaningful. For example, by penalizing candidate models that violate mass conservation or that predict negative concentrations, the search space is refined to favor models that adhere to known prior knowledge. This constraint-based approach not only improves the predictive reliability of the resulting models but it also reduces the experimental cost for their discovery, which is especially important when experiments are expensive or difficult to run.

The second enhancement is the incorporation of uncertainty quantification in the model predictions. While point estimates provide a single best-fit model, they do not offer insight into the confidence or reliability of the predictions. By adopting a sampling-based uncertainty quantification method, such as the Metropolis-Hastings algorithm, we are able to generate a posterior distribution over the kinetic parameters. This probabilistic framework enables us to assess the variability of the model outputs and to estimate confidence intervals for predictions. The ability to quantify uncertainty is particularly important for decision-making in safety-critical applications, where understanding the range of possible outcomes can guide more informed process control and risk management strategies. 

Together, these enhancements address some of the key challenges that have limited the broader adoption of automated kinetic modeling methods. The introduction of mathematical constraints effectively narrows the search space of the genetic programming algorithm, which focuses the computational resources and mitigates the risk of converging to physically implausible models. Simultaneously, uncertainty quantification provides a robust mechanism to evaluate the reliability of the selected models, ensuring that their predictions are accompanied by meaningful measures of confidence. As a result, the extended framework not only reduces the experimental burden by efficiently guiding the model discovery process, but also significantly improves the trustworthiness of the discovered kinetic models.

In summary, the need for reliable and interpretable kinetic models is critical in the advancement of catalytic process engineering. By merging the strengths of symbolic regression-based automated knowledge discovery frameworks with the rigor of physics-based constraints and uncertainty quantification, the extended framework presented in this paper represents a step forward in the field. Not only does it offer a more physically consistent and robust approach to model discovery, but it also provides the necessary tools to assess prediction reliability: a feature that is indispensable for the safe and efficient design of chemical processes. This work thus opens new avenues for the development of automated kinetic models that are both data-efficient and deeply rooted in physical principles.

The remainder of the paper is organized as follows. In Section \ref{Methodology}, we first describe the underlying automated knowledge discovery framework that forms the foundation of our work. We then detail how physical constraints were integrated into the genetic programming algorithm, drawing on past findings regarding constraint inclusion, and explain our approach to quantifying the uncertainty of model predictions. Section \ref{Case Studies} outlines the selection and rationale behind the case studies used to evaluate our new framework, Physics Informed Automated Discovery of Kinetics (PI-ADoK), which is benchmarked against the original ADoK version. Section \ref{Results and Discussions} presents the computational results and discusses their implications, and finally, Section \ref{Conclusions} concludes the paper with a summary of our main contributions and suggestions for future work.


\section{Methodology}\label{Methodology}
We begin by outlining our methodology, termed PI-ADoK (Physics Informed Automated Discovery of Kinetics). The framework operates through three main phases. First, we use a genetic programming algorithm guided by both data and domain knowledge (in the form of physical constraints) to generate candidate models that are consistent with known chemical principles. Second, we apply a sequential optimization routine to accurately estimate the parameters of these promising candidates. Finally, we employ a transparent model selection procedure based on the Akaike Information Criterion (AIC) to identify the best model. We opted for an information criterion rather than a data-splitting strategy because it allows the full dataset to be used for model construction while still providing a rigorous evaluation mechanism: an approach that is particularly advantageous when data are scarce. Our choice to use AIC, as opposed to any other criterion, can be found in the `Supplementary Information' of \citet{deCarvalhoServia_2024}.

PI-ADoK adopts a conventional symbolic regression approach, often referred to as the strong formulation \cite{Bertsimas_2023}, which relies on rate measurements to derive kinetic models. Because these rates are not directly measured in experiments, they must be approximated. Following our three-phase process, the framework first determines optimal concentration profiles that describe how species concentrations evolve over time. These profiles are then numerically differentiated to estimate the reaction rates. With these estimates, the same three-step process is repeated to identify the kinetic rate model that best describes the observed behavior. The resulting model is integrated and its predictions compared with the original concentration data.

Our genetic programming-based strategy for estimating rates has demonstrated superior performance compared to many state-of-the-art methods (as detailed by \citet{VanBreugel_2020}), with further details provided in the `Supplementary Information' of \citet{deCarvalhoServia_2024}. It is important to note that the time-series kinetic data needed to implement PI-ADoK can be acquired either from transient experiments, where the evolution of species concentrations is monitored over time in batch reactors, or from steady-state experiments, which measure concentrations as a function of residence time in plug-flow reactors.

\begin{figure}[!htb]
    \centering
    \includegraphics[width=1\textwidth]{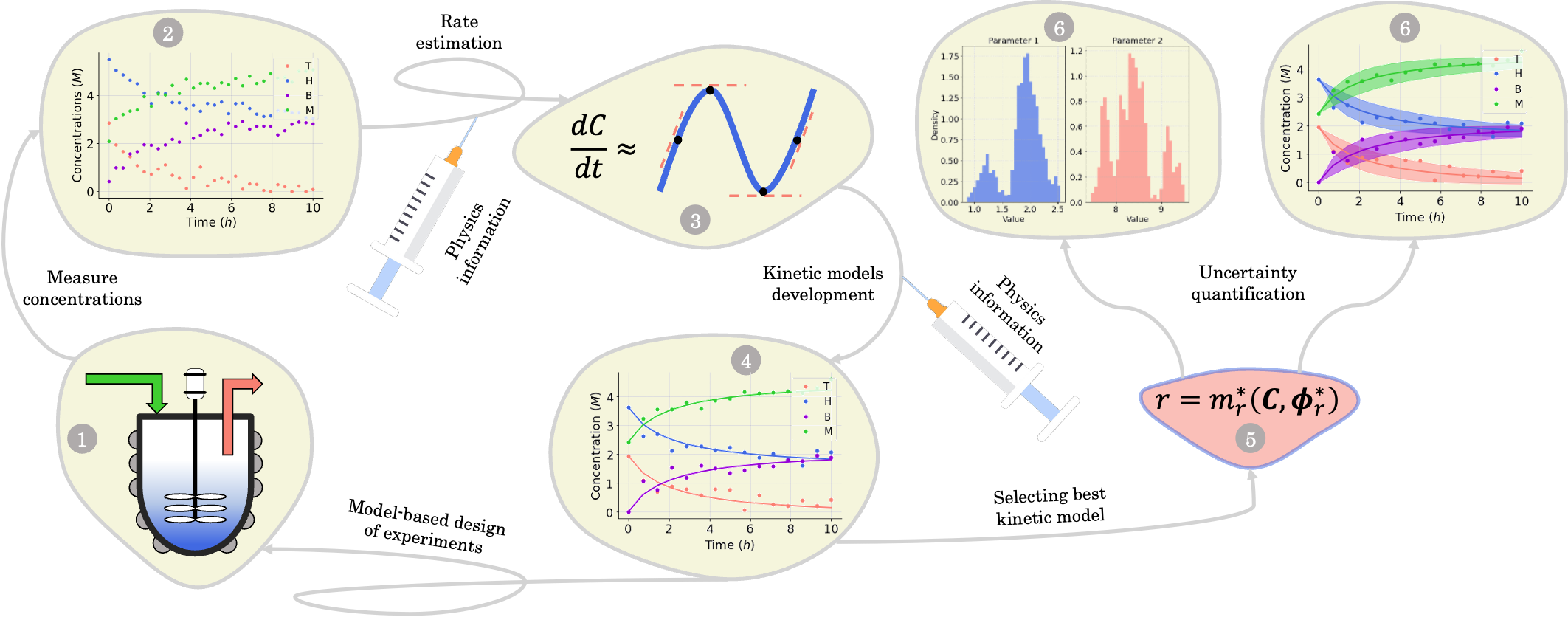}
    \caption{Conceptual overview of the Physics Informed Automated Discovery of Kinetics (PI-ADoK) framework. Experimental data are first collected from an experiment/simulation and used to construct symbolic concentration models that are injected with prior knowledge. Numerical differentiation of these models provides estimated reaction rates, which guide the discovery of a kinetic rate expression that is also injected with prior knowledge. If the final model is satisfactory, the uncertainty of its prediction can be quantified. This cycle may be iterated with additional experiments, informed by model-based design of experiments, until a satisfactory model is found.}
    \label{Fig:high_level_diagram}
\end{figure}

The methodology is designed as a closed-loop system. If the initial model output is unsatisfactory due either to deviations from established physical principles (for example, neglecting the influence of a species believed to affect the reaction rate) or due to inadequate fitting of the kinetic non-linearities, the modeler can initiate an optimal experiment tailored for the specific discovery task, as determined by model-based design of experiments (MBDoE). The new experimental data can then be merged with the original dataset, and the entire process iterated until a satisfactory model is obtained or the experimental budget is exhausted. Additionally, these targeted experiments could also serve to validate the accuracy of previously proposed models alongside the AIC-based selection. Once the iterative process concludes, the uncertainty of the final, best candidate model is quantified, providing insight into the reliability of its predictions. A high-level diagram of the PI-ADoK workflow is presented in Fig. \ref{Fig:high_level_diagram}, with further details available in Fig. \ref{Fig:pi_adok_flowchart}.

In developing this methodology, we intentionally chose a genetic programming-based approach (even though it may sometimes diverge from traditional mass-action laws) in favor of an automated strategy that requires a priori specification of potential chemical reactions involved in the reactive system being investigated. This choice is justified by several advantages. First, our approach eliminates the need to assume predefined reaction families or perform extensive thermodynamic calculations, both of which can be prohibitive due to their complexity or lack of available data. Second, it is designed to extract essential kinetic information in contexts where prior knowledge is minimal or absent. Third, the method retains the flexibility to incorporate expert knowledge through mathematical constraints whenever available, thereby aligning the discovered models with established physical phenomena. In essence, our methodology is well-suited to handle scenarios with limited prior information while effectively utilizing any available knowledge, making it a robust and versatile tool for kinetic model discovery in chemical systems.

We begin by establishing the mathematical notation necessary to precisely describe our methodology. First, we adopt the standard symbolic regression formulation \cite{Virgolin_2022}, which serves as the foundation before introducing the strong formulation of our approach.

Let the set \(\mathcal{Z}\) be defined as the union of an arbitrary collection of constants, \(\Gamma\), and a fixed set of variables, \(\mathcal{X}\). The operator set \(\mathcal{P}\) consists of both arithmetic operations (\(\diamond: \mathbb{R}^n \rightarrow \mathbb{R}\)) and a finite collection of special one-dimensional functions (\(\Lambda: \mathbb{R} \rightarrow \mathbb{R}\)). Through iterative function composition using the operators in \(\mathcal{P}\) over the elements in \(\mathcal{Z}\), we form the model search space \(\mathcal{M}\).

In our framework, variables are represented as state vectors \(x \in \mathbb{R}^{n_x}\). Each data point comprises a state \(x\) and its corresponding target value \(y \in \mathbb{R}\) generated by an unknown function \(f: \mathbb{R}^{n_x} \rightarrow \mathbb{R}\), such that \(y = f(x)\). Collectively, the dataset is given by \(\mathcal{D} = \left\{ \left( x^{(i)}, y^{(i)} \right) \mid i = 1, \ldots, n_t \right\}\). To measure the discrepancy between predictions and target values, we employ a suitable positive-valued function \(\ell: \mathbb{R}^{n} \times \mathbb{R}^{n} \rightarrow \mathbb{R}^{+}\).

A symbolic model \(m \in \mathcal{M}\) is characterized by a finite set of parameters \(\theta_m\), whose dimensionality \(d_m\) depends on the specific model. We denote the model's prediction under parameters \(\theta_m\) as \(m(\cdot \mid \theta_m)\), and we represent the predicted value by \(\hat{y}_m\) (i.e., \(\hat{y}_m = m(\cdot \mid \theta_m)\)). Crucially, our approach has two phases: the first where the main objective is to find the optimal model structure, and the second where the main objective is to fine-tune the optimal model structure and discover its optimal parameters. We define the optimal model \(m^*\) as the model that minimizes the sum of the data fitting error and a penalty term proportional to the degree of constraint violation. Formally, this is expressed as:

\begin{equation} \label{eq:SR}
    m^* = \argmin_{m \in \mathcal{M}} \left\{ \sum_{i=1}^{n_t} \ell\left( \hat{y}_m^{(i)}, y^{(i)} \right) + \sum_{j=1}^{J} \lambda_j \, P_j(m) \right\},
\end{equation}

where \(P_j(m)\) quantifies the violation of the \(j\)-th constraint, \(\lambda_j\) is a constant scaling factor specific to that constraint, and \(J\) is the total number of constraints.

The corresponding optimal parameters are determined by

\begin{equation} \label{eq:PE}
    \theta_{m^*}^* = \argmin_{\theta_{m^*}} \left\{\sum_{i=1}^{n_t} \ell\left( \hat{y}_{m^*}^{(i)}, y^{(i)} \right) + \sum_{j=1}^{J} \lambda_j \, P_j(m)\right\}.
\end{equation}

In the context of dynamical systems, the state variables are functions of time, \(x(t) \in \mathbb{R}^{n_x}\), representing the evolution of the system over a fixed interval \(\Delta t = [t_0, t_f]\). The system dynamics are characterized by the time derivatives \(\dot{x}(t) \in \mathbb{R}^{n_x}\) and the initial condition \(x_0 = x(t_0)\).

For our kinetic rate models, we assume that the \(n_t\) sampling times \(t^{(i)}\) lie within the interval \(\Delta t\). The concentration measurements \(C\) at each time \(t^{(i)}\) approximate the true state \(x(t^{(i)})\), while the rate estimates \(r\) approximate the corresponding time derivatives, \(r^{(i)} \approx \dot{x}(t^{(i)})\). Thus, the dataset becomes \(\mathcal{D} = \left\{ \left( t^{(i)}, C^{(i)} \right) \mid i = 1, \ldots, n_t \right\}.\)

As before, we denote model predictions by a hat: \(\hat{C}_m\) for states and \(\hat{r}_m\) for rates, with the outputs given by \(\hat{C}_m(\cdot \mid \theta_m)\) and \(\hat{r}_m(\cdot \mid \theta_m)\), respectively.

We quantify the complexity of a model using the function \(\mathcal{C}(m)\), here defined as the number of nodes in the expression tree representing the model \cite{Cranmer_2023}. Models can then be grouped into families based on their complexity level \(\kappa \in \mathbb{N}\), denoted as \(\mathcal{M}^\kappa = \left\{ m \in \mathcal{M} \mid \mathcal{C}(m) = \kappa \right\}.\)

This notation establishes the mathematical foundation for our methodology, facilitating a clear and systematic description of our approach to automated kinetic model discovery.

\subsection{Introduction to the Strong Formulation}\label{Strong formulation}
Before getting into the detailed explanations of model generation, model selection, mathematical constraints, and uncertainty quantification, we first provide a concise, itemised workflow of PI-ADoK. This overview will serve as a road-map for the discussion that follows.

\begin{enumerate}
  \item \textbf{Data collection:} Acquire time–series concentrations $\!\bigl(t,\;C_i(t)\bigr)$ of all reactants and products.
  \item \textbf{Generate constrained concentration surrogates:} Employ genetic programming with embedded physical constraints (positivity, equilibrium,\,\dots) to build differentiable symbolic models $\eta_i(t)$ that fit the measured $C_i(t)$.
  \item \textbf{Parameter refinement (concentration):} Calibrate every surrogate by solving Eq.~\eqref{eq:PE} to obtain $\theta_{\eta_i}^{\star}$.
  \item \textbf{Model selection (concentration):} Use $\mathrm{AIC}$ to pick the most accurate yet parsimonious $\eta_i(t)$ from the model set for each chemical species in each experiment.
  \item \textbf{Derivative estimation:} Differentiate the chosen $\eta_i(t)$; the derivatives $\dot{\eta}_i(t)$ provide rate estimates $r_i(t)$.
  \item \textbf{Generate constrained rate model candidates:} Apply genetic programming with constraints to the rate data, yielding a set $\mathcal M^{\kappa}$ of symbolic rate models for each complexity $\kappa$.
  \item \textbf{Parameter refinement (rates):} Optimize every rate model by solving the inner problem in Eq.~\eqref{eq:PE Strong}.
  \item \textbf{Model selection (rates):} Rank the $\kappa$-winners with $\mathrm{AIC}$ and select the final kinetic expression $m^{\star}$.
  \item \textbf{Optional MBDoE loop:} If $m^{\star}$ is unsatisfactory and budget remains, use model-based design of experiments to propose new conditions (default: discriminate between the best and second-best rate models), collect data, and return to Step 2.
  \item \textbf{Uncertainty quantification:} For the accepted model, quantify parameter uncertainty (with Metropolis–Hastings) and propagate it to obtain predictive intervals.
\end{enumerate}

For PI-ADoK, which leverages the strong formulation of symbolic regression, the primary objective is to determine the model \(m\) that best maps the state variables \(x(t)\) to the corresponding rates $r^{(i)}$, i.e.,

\begin{equation} \label{eq:strog_model}
    \hat{r}_m(t \mid \theta_m) = m(x(t) \mid \theta_m).
\end{equation}

Since direct measurements of the rates \(r^{(i)}\) are unavailable, they must first be estimated from the concentration data \(C^{(i)}\). To this end, our approach constructs an intermediate symbolic model \(\eta\) that approximates the concentration measurements, such that \(\eta(t^{(i)}) \approx C^{(i)}\). This process follows the standard symbolic regression procedure, as described in Eqs. \eqref{eq:SR} and \eqref{eq:PE}, with the associated model selection methodology detailed in Section \ref{Model Selection}.

Overfitting is inherently controlled at two distinct stages of the PI-ADoK workflow. First, during the genetic programming search, the population is arranged by structural complexity $\kappa$. For every admissible dimensionality (e.g.\ $\kappa = 3,4,5,\ldots$) the algorithm independently seeks and stores the best performing model before any cross-complexity comparison is made. This level-wise competition ensures that simple models are never forced to compete directly with much richer expressions and by defining an upper limit of complexity, the search process is prevented from drifting toward unnecessarily intricate solutions. Second, when the set of level-wise winners is compared to choose the final model, we employ the Akaike Information Criterion, which adds an explicit penalty that grows with the dimensionality of the model. By coupling complexity-arranged search with AIC-based selection, PI-ADoK guards against overfitting both during model generation and during the ultimate selection of the governing kinetic expression.

Because the model \(\eta\) is differentiable, its derivative, \(\dot{\eta}(t^{(i)})\), serves as an approximation for the true rates, i.e., \(\dot{\eta}(t^{(i)}) \approx r^{(i)}\). With these rate estimates in hand, we can formulate the optimization problem as follows. At the outer level, we optimize over candidate models of fixed complexity \(\kappa\) by minimizing the sum of the fitting error and a penalty term that is proportional to the degree of constraint violation:

\begin{equation}\label{Eq:1.1}
    m^\star = \argmin_{m \in \mathcal{M}^\kappa} \left\{ \sum_{i=1}^{n_t} \ell \left(\hat{r}_m(t^{(i)} \mid \theta_m), r^{(i)}\right) + \sum_{j=1}^{J} \lambda_j \, P_j(m) \right\}.
\end{equation}

At the inner level, we optimize the parameters of the selected model \(m^\star\) as follows:

\begin{equation}\label{eq:PE Strong}
    \theta_{m^\star}^\star = \argmin_{\theta_{m^\star}} \left\{\sum_{i=1}^{n_t} \ell \left(\hat{r}_{m^\star}(t^{(i)} \mid \theta_{m^\star}), r^{(i)}\right) + \sum_{j=1}^{J} \lambda_j \, P_j(m) \right\}.
\end{equation}

\begin{figure}[!htb]
    \centering
    \includegraphics[width=1\textwidth]{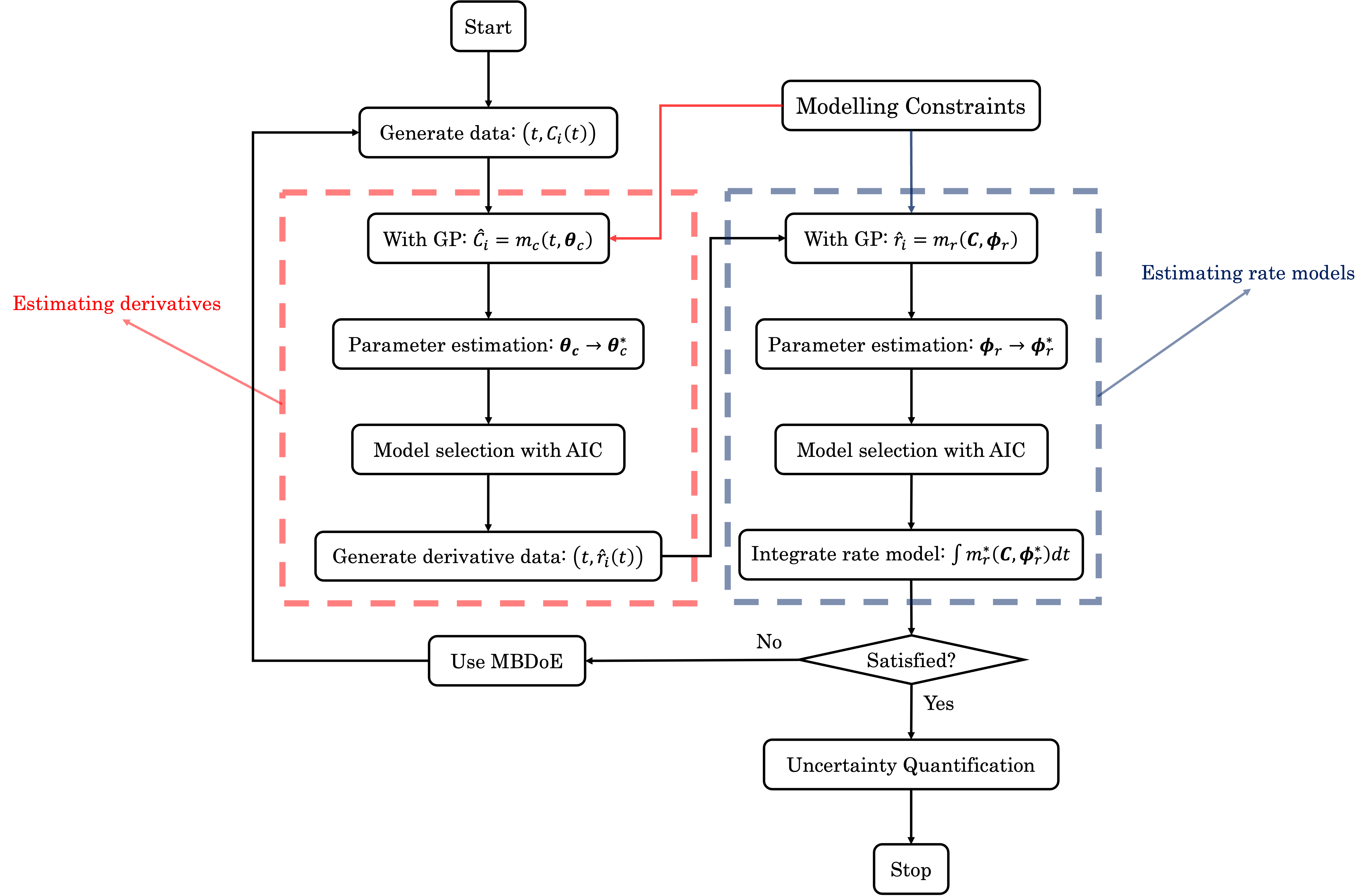}
    \caption{Step-by-step flow of PI-ADoK, highlighting the two main tasks: estimating derivatives (red box) and generating rate models (blue box). In the derivative-estimation phase, genetic programming produces candidate concentration models, followed by parameter estimation and model selection via AIC. These models are then numerically differentiated to approximate reaction rates. In the rate-modeling phase, the framework uses the estimated rates to build kinetic expressions, again refining candidates through parameter estimation and model selection. Model-based design of experiments (MBDoE) can propose new experiments to collect data if the current model is unsatisfactory, closing the loop until a reliable model is obtained. Uncertainty quantification is then performed on the final model to assess prediction reliability. Constraints are included in each step of model construction to guide the genetic programming algorithm to physically-sensible models.}
    \label{Fig:pi_adok_flowchart}
\end{figure}

In both Eqs. \eqref{Eq:1.1} and \eqref{eq:PE Strong}, the function \(\ell\) represents the sum of squared errors (SSE). The Limited-memory Broyden-Fletcher-Goldfarb-Shanno (L-BFGS) algorithm is employed for solving the parameter estimation problem \cite{Liu_1989}. L-BFGS is well-suited for handling this problem due to its performance in tasks pertaining to parameter estimation and optimization \cite{Malouf_2002, Liu_1989}. The stopping criteria for the optimization are left to the default options in the Scipy package \cite{2020SciPy-NMeth}, and a multi-start approach is employed, where multiple runs are initiated with different starting points, and the best solution is retained. A schematic overview of the complete PI-ADoK workflow is presented in Fig. \ref{Fig:pi_adok_flowchart}.

The PI-ADoK framework is designed to handle complex chemical reaction scenarios, including cases with multiple reactions occurring in parallel or sequentially. In this work, however, we focus on single-reaction systems. For multi-reaction systems, the approach is significantly different. Instead of deriving a single unified model to describe the kinetic rates of all species, the chemical system would require PI-ADoK to develop individual models for each reactant and product. This is due to the fact that, in multi-reaction systems, the dynamics of each species are governed by distinct mathematical functions, with no direct stoichiometric relationships linking their rates. An example of applying the strong formulation of symbolic regression to multi-reaction systems is provided in the `Supplementary Information' of \citet{deCarvalhoServia_2024}.

\subsection{Model Selection}\label{Model Selection}
Having outlined in Section \ref{Strong formulation} how PI-ADoK produces a level-wise set of candidate models (one best expression for every structural complexity $\kappa$) we now turn to the question of how to choose among those winners.  The selection step must favor models that are predictive yet parsimonious, thereby reinforcing the overfitting defenses already built into the search procedure.

Instead of employing a data-splitting approach for model selection, PI-ADoK leverages an information criterion, allowing the entire dataset to be utilized for both model construction and evaluation. This is particularly beneficial in low-data environments, as it maximizes the amount of information available for identifying suitable kinetic models.

We specifically adopt the Akaike Information Criterion (AIC) based on prior comparative analyses of different information criteria, where AIC consistently demonstrated superior performance in kinetic discovery \cite{deCarvalhoServia_2023}. Formally, for a model \(m\) with parameter set \(\theta_m\) of dimension \(d_m\), the AIC is given by:

\begin{equation}\label{eq:AIC}
    \text{AIC}_m = 2 \, NLL\bigl(\theta_m \mid \mathcal{D}\bigr) + 2\,d_m,
\end{equation}

where \(NLL\) denotes the negative log-likelihood \cite{Akaike_1974}. When comparing two models \(m_1\) and \(m_2\), the one with the lower AIC value from Eq.~\eqref{eq:AIC} is deemed preferable.

\subsection{Model-Based Design of Experiments}
If the dataset used for model discovery is insufficient to yield an adequate model, and provided the experimental budget has not been exhausted, we can leverage insights from the optimized models to design a more informative experiment. In particular, we identify the operating conditions that maximize the discrepancy between the state predictions $\hat x(t|\theta^\star)$ of the two best proposed models, denoted as $\eta$ and $\mu$, based on the current dataset. The rationale for selecting these two models is discussed in \citet{deCarvalhoServia_2024}. The MBDoE approach adopted in this work follows the framework developed by \citet{Hunter_1965}:

\begin{equation}\label{Eq: MBDoE}
     x_0^{(new)} = \argmax_{x_0} \left\{ x_0 + \int_{t_0}^{t_f} \ell\left(\hat x_\eta \left(\tau\mid\theta_\eta^\star \right), \hat x_\mu \left(\tau\mid\theta_\mu^\star \right) \right)\, d\tau \right\}.
\end{equation}

In Eq. \eqref{Eq: MBDoE}, $\ell$ represents the SSE. Once the optimal initial conditions are determined, a new experiment can be performed to generate additional data points, which are then incorporated into the original dataset. With this enriched dataset, PI-ADoK can be executed again, thereby closing the loop between informative experimental design and optimal model discovery.

\subsection{Integration of Mathematical Constraints}
The incorporation of mathematical constraints into symbolic regression frameworks has attracted considerable attention in the literature, yielding mixed outcomes. On one hand, studies such as those by \citet{Kronberger_2022} indicate that integrating constraints may lead to higher prediction errors on both training and testing datasets. They attribute this effect to slower convergence rates and a more rapid loss of genetic diversity. Nevertheless, this same study suggest that under elevated noise levels (which often mirror the inherent variability in experimental setups) the benefits of enforcing constraints become more pronounced by steering the search toward models that are consistent with the underlying system.

Further investigations by \citet{Haider_2023} extended these observations by examining case studies under conditions of high noise. Their findings indicate that, although the improvements in prediction error were sometimes not statistically significant compared to unconstrained approaches, the incorporation of constraints did help in identifying models with a lower propensity for overfitting and enhanced adherence to expected behavior. In addition, research by \citet{Błądek_2019} demonstrates that for smaller datasets (typical of many experimental scenarios) the integration of mathematical constraints can yield statistically significant improvements over traditional genetic programming (GP) algorithms without constraints.

Taken together, these studies, despite their ambiguous outcomes, are encouraging for our application area. Experimental data are frequently characterized by high noise levels and limited sample sizes, conditions under which the selective enforcement of constraints appears to offer tangible benefits. This suggests that, even if the addition of constraints occasionally incurs a trade-off in prediction accuracy, the overall improvements in physical plausibility and model robustness make this approach a promising avenue for experimental applications like the one we deal with in this work.

Motivated by these findings there is a clear need for a flexible methodology to incorporate extensive prior knowledge (often available in kinetic studies) into GP. PI-ADoK integrates constraints directly into the GP process to ensure that candidate models not only fit the data but also conform to established physical laws.

Integrating constraints into GP is a delicate endeavor that requires balancing exploration and exploitation in a vast search space. On one hand, constraints reduce the search space by eliminating models that violate known physical principles, thus focusing computational effort on promising regions. On the other hand, overly stringent constraints lead to reduced population diversity, which can induce premature convergence, and inevitably results in suboptimal solutions.

In PI-ADoK, constraints are incorporated in a straightforward yet effective manner. Each candidate model is evaluated based on its prediction error and its compliance with a set of predefined constraints. Specifically, our constraints verify that candidate models:


\begin{enumerate}
    \item Exactly respect the initial conditions (since these are determined with minimal uncertainty).
    \item Reach equilibrium so that the function's end behavior converges to a constant value.
    \item Consistently predict outputs with the correct sign (e.g., positive concentrations or negative rates).
    \item Exhibit the correct monotonic behavior, being either always increasing or always decreasing.
\end{enumerate}

Each of these constraints can be turned on and off independently based on the chemical system being investigated. When a candidate model satisfies all constraints, its fitness is determined solely by its prediction error. However, if it violates one or more constraints, a penalty, which is proportional to the degree of violation and scaled by a user-defined hyperparameter, is added to its fitness. This penalty-based method enables fine-tuning of the balance between allowing some flexibility in the search and enforcing strict constraint adherence through the hyperparameters. It is important to note that these hyperparameters were manually fine-tuned for our experiments. Although a more formal hyperparameter optimization could potentially enhance the robustness of our findings, we believe that these parameters should be tuned on a case-by-case basis, since the appropriate confidence in the constraints depends on the specific system, the amount of available information, and ultimately the performance of the algorithm.

This approach offers several advantages:

\begin{itemize}
    \item It preserves the interpretability and physical plausibility of the resulting models by ensuring adherence to known physical laws.
    \item It focuses the search on promising regions of the model space, potentially reducing the experimental cost of model discovery.
    \item The use of hyperparameters to scale penalty terms allows the algorithm to be tailored to different problem contexts, balancing the need for exploration with the drive for exploitation.
\end{itemize}

However, it is important to note that our current implementation employs static hyperparameters that remain constant throughout the search process. In future work, it would be worthwhile to investigate dynamic hyperparameter tuning strategies, where the penalty factors evolve during the search. For instance, one might hypothesize that a more relaxed constraint regime in the early stages could maximize diversity and facilitate a broad exploration of the model space. As the search progresses and promising regions are identified, the constraints could gradually become more stringent, thereby focusing computational resources on refining high-performing solutions.

\subsection{Uncertainty Quantification Using the Metropolis-Hastings Algorithm}

Uncertainty quantification is an important aspect of modeling complex kinetic systems, as it provides insight into the confidence and robustness of predicted model behavior. In the context of symbolic regression, and specifically for PI-ADoK, the need to accurately propagate uncertainty through non-linear, high-dimensional kinetic models have led us to adopt a sampling-based approach using the Metropolis-Hastings (MH) algorithm.

Various methods exist for uncertainty quantification, ranging from simpler techniques such as Laplace approximations and sigma points to more sophisticated sampling algorithms like Hamiltonian Monte Carlo (HMC) and MH. For our purposes of kinetic modeling, where accuracy may be critical, the MH algorithm was selected because of its ability to handle complex, non-linear distributions whilst having a simple and intuitive implementation that provides effective results. This flexibility in choosing proposal distributions makes MH particularly adaptable to the intricate dynamics often encountered in kinetic modeling.

The MH algorithm is an iterative method designed to sample from a target distribution: in our case, the posterior distribution of the model parameters. It works by constructing a Markov chain, meaning that each new sample depends only on the current state, and as the chain evolves, its distribution converges to the target distribution (this convergence is known as the chain reaching its stationary distribution).

At each iteration, a candidate point is generated by perturbing the current point using a proposal distribution. The candidate is then either accepted or rejected based on an acceptance probability. This probability is calculated to satisfy the detailed balance condition, which essentially ensures that the likelihood of moving from one point to another and vice versa is balanced in such a way that the chain will eventually reflect the target distribution.

In our implementation, if the candidate improves the model's fit (i.e., it has a higher posterior probability) or meets the acceptance criterion probabilistically even when it is less likely than the current state, the candidate is accepted and becomes the new current state. If not, the algorithm retains the current state. This process of generating, evaluating, and either accepting or rejecting candidates allows the chain to explore the parameter space effectively. Over many iterations, the samples collected approximate the posterior distribution, providing a robust quantification of uncertainty in our kinetic models.

The main steps of the MH algorithm are summarized in Algorithm~\ref{alg:MH_inference}.

\begin{algorithm}[htb!]
\caption{Metropolis-Hastings Algorithm for Kinetic Parameter Inference}
\label{alg:MH_inference}
\begin{algorithmic}[1]
\Require Initial parameters \(\theta_0\) (a non-negative vector); number of iterations \(N\); Gaussian distribution with standard deviation \(\sigma\) (i.e., \(q(\theta' \mid \theta)=\mathcal{N}(\theta, \sigma^2)\)).
\Ensure A sequence of parameter samples \(\{\theta_0, \theta_1, \ldots, \theta_N\}\) approximating the posterior distribution \(p(\theta \mid \mathcal{D})\).
\State \textbf{Define} the likelihood function: 
\[
L(\theta)=\exp\left(-\frac{\mathrm{SSE}(\theta)}{2}\right),
\]
where \(\mathrm{SSE}(\theta)\) is the sum of squared errors from the kinetic model.
\State \textbf{Define} the prior density \(p_{\text{prior}}(\theta)\). (For this study, the prior density is a multivariate normal with specified mean and covariance. The specified mean is defined by the result obtained by solving Eq. \eqref{eq:PE Strong} for the chosen model, and the specified covariance is defined based on our level of confidence of our defined mean. These design choices were made so that moderately informative priors, which are usually available in kinetic studies, can be directly introduced in the framework.)
\State \textbf{Define} the unnormalized target (posterior) density:
\[
p(\theta) \propto L(\theta) \cdot p_{\text{prior}}(\theta).
\]
\State Set \(\theta \gets \theta_0\).
\State Initialize the sample set \(\mathcal{S} \gets [\,]\).
\For{\(i = 1\) to \(N\)}
    \State Generate a candidate \(\theta' \sim \mathcal{N}(\theta, \sigma^2)\) 
    \State Enforce non-negativity: \(\theta' \gets \max(\theta', 0)\) 
    \State Compute the current target: \(P_{\text{current}} = L(\theta) \cdot p_{\text{prior}}(\theta)\).
    \State Compute the proposed target: \(P_{\text{proposed}} = L(\theta') \cdot p_{\text{prior}}(\theta')\).
    \State Calculate the acceptance probability:
    \[
    a = \min\left\{ 1, \frac{P_{\text{proposed}}}{P_{\text{current}}} \right\}.
    \]
    \State Draw \(u \sim \text{Uniform}(0,1)\).
    \If{\(u < a\)}
        \State Set \(\theta \gets \theta'\).
    \Else
        \State Retain \(\theta\).
    \EndIf
    \State Append the current \(\theta\) to \(\mathcal{S}\).
\EndFor
\State \Return \(\mathcal{S}\).
\end{algorithmic}
\end{algorithm}


A key advantage of the MH algorithm is its capability to propagate uncertainty through the model in a robust manner. By drawing samples from the posterior distribution, we can estimate credible intervals and other summary statistics that characterize the uncertainty associated with model predictions. Despite its computational intensity and the need for careful tuning of the proposal distribution, MH remains one of the most robust methods available for uncertainty quantification in complex systems.

Our implementation uses a candidate-generating density that is carefully chosen to balance the trade-off between exploration and computational efficiency. The proposal distribution parameters were adjusted experimentally to achieve an acceptance rate in the range of 40\% to 50\%, which we found to be optimal for our kinetic models. In doing so, the MH algorithm is able to sample effectively from regions of the parameter space that contribute most to predictive uncertainty.
 
When implementing the MH algorithm for uncertainty quantification, several practical issues must be addressed. First, the choice of the proposal distribution is crucial; it must be sufficiently broad to explore the parameter space, yet not so broad that the acceptance rate becomes prohibitively low. Second, the convergence of the Markov chain must be carefully monitored, typically using diagnostic tools such as autocorrelation analysis or the Gelman-Rubin statistic, to ensure that the sampled values are representative of the target distribution. In our experiments, we discard an initial set of samples (the burn-in period) to mitigate the influence of the starting point, and then collect a large number of samples to reliably estimate the posterior distribution.

While our current work demonstrates the feasibility of using the MH algorithm for uncertainty quantification in kinetic models, several avenues for future research remain. For instance, comparing MH with alternative sampling methods like HMC may yield insights into strategies that balance computational efficiency and accuracy differently.

In summary, the use of the MH algorithm in our framework enables robust uncertainty quantification by effectively sampling from the posterior distribution of kinetic model parameters. Despite challenges such as increased computational cost and the need for meticulous tuning, MH provides a powerful tool for capturing the inherent uncertainty in model predictions.

\section{Catalytic Kinetic Case Studies}\label{Case Studies}
To evaluate the performance of our extended framework, PI-ADoK, we compared it against its original counterpart, ADoK-S, using three catalytic reaction case studies drawn from the literature. The selected case studies encompass a variety of kinetic complexities, from the relatively straightforward isomerization reaction to the more complex hydrodealkylation of toluene. This diversity ensures that our framework is tested across a wide spectrum of reaction types and data conditions, resembling the kinds of datasets typically obtained from experimental setups. For conciseness, our discussion focuses primarily on one of the examples -- the decomposition of nitrous oxide.

By comparing PI-ADoK with its original version, ADoK-S, across these case studies, we aim to demonstrate that our extended framework is capable of producing models that not only fit the observed data but also adhere to expected physical behavior whilst minimizing the experimental cost. This focus on accuracy, physical plausibility and resource optimization is crucial for developing reliable and cost-effective kinetic models in chemical engineering. 

\subsection{The Decomposition of Nitrous Oxide}\label{Decomposition}
The decomposition of nitrous oxide is modeled by:

\begin{equation}\label{eq:2.3}
    2N_2O \rightleftharpoons 2N_2 + O_2,
\end{equation}

with the reaction rate expressed as:

\begin{equation}\label{eq:2.4}
    r = -2\frac{dC_{N_2O}}{dt} = 2\frac{dC_{N_2}}{dt} = \frac{dC_{O_2}}{dt} = \frac{k_A\,C_{N_2O}^2}{1 + k_B\,C_{N_2O}},
\end{equation}

where the parameters are set as \(k_A = 2\) M\(^{-1}\) h\(^{-1}\) and \(k_B = 5\) M\(^{-1}\) \cite{Levenspiel_1998}. An in-silico dataset is generated with \(\Delta t = [0, 10]\) h and \(n_t = 15\) samples, based on five experiments with initial conditions selected from a \(2^k\) factorial design: \((C_{N_2O}(0), C_{N_2}(0), C_{O_2}(0)) \in \{(5, 0, 0),\ (10, 0, 0),\ (5, 2, 0),\ (5, 0, 3),\ (0, 2, 3)\}\). 

For all experiments, the system is assumed to be isochoric and isothermal, and Gaussian noise with zero mean and a standard deviation of 0.2 is added to each measurement to simulate realistic experimental conditions. Figure~\ref{Fig:pi_adok_results} a) and \ref{Fig:pi_adok_results} e) illustrate the dataset for two of the experiments. The total of 75 data points per case reflects a realistic experimental scenario, particularly in light of advancements in high-throughput kinetic studies \cite{Schrecker_2023, Waldron_2020, Taylor_2021}.

\subsection{The Hydrodealkylation of Toluene}\label{Hydrodealkylation}
The hydrodealkylation of toluene reaction is represented by:

\begin{gather}\label{Eq:2.1}
    C_6H_5CH_3 + H_2 \rightleftharpoons C_6H_6 + CH_4,
\end{gather}

with the corresponding rate expression given by:

\begin{gather}\label{Eq:2.2}
    r = -\frac{dC_T}{dt} = -\frac{dC_H}{dt} = \frac{dC_B}{dt} = \frac{dC_M}{dt} = \frac{k_A\,C_T\,C_H}{1 + K_B\,C_B + K_C\,C_T},
\end{gather}

where \(C_T\), \(C_H\), \(C_B\), and \(C_M\) denote the concentrations of toluene, hydrogen, benzene, and methane, respectively. The kinetic parameters are defined as \(k_A = 2\) M\(^{-1}\) h\(^{-1}\), \(K_B = 9\) M\(^{-1}\), and \(K_C = 5\) M\(^{-1}\) \cite{Fogler_2016}. 

Based on Eq. \eqref{Eq:2.2}, we generated an in-silico dataset over a time interval \(\Delta t = [0, 10]\) h with \(n_t = 15\) sampling points. Five experiments were simulated with different initial conditions, chosen randomly from a \(2^k\) factorial design: \((C_T(0), C_H(0), C_B(0), C_M(0)) \in \{(1, 8, 2, 3),\ (5, 8, 0, 0.5),\ (5, 3, 0, 0.5),\ (1, 3, 0, 3),\ (1, 8, 2, 0.5)\}\). Gaussian noise (zero mean, standard deviation 0.2) is added to mimic measurement uncertainties.

\subsection{The Theoretical Isomerization Reaction}\label{Isomerization}
The isomerization reaction is described by:

\begin{equation}\label{eq:2.5}
    A \rightleftharpoons B,
\end{equation}

with the kinetic rate given by:

\begin{equation}\label{eq:2.6}
    r = -\frac{dC_A}{dt} = \frac{dC_B}{dt} = \frac{k_A\,C_A - k_B\,C_B}{k_C\,C_A + k_D\,C_B + k_E},
\end{equation}

where the rate constants are \(k_A = 7\) M h\(^{-2}\), \(k_B = 3\) M h\(^{-2}\), \(k_C = 4\) h\(^{-1}\), \(k_D = 2\) h\(^{-1}\), and \(k_E = 6\) M h\(^{-1}\) \cite{Marin_2019}. An in-silico dataset is generated with \(\Delta t = [0, 10]\) h and \(n_t = 15\) data points, using five experiments with initial conditions drawn from a \(3^k\) factorial design: \((C_A(0), C_B(0)) \in \{(2, 0),\ (10, 0),\ (2, 2),\ (10, 2),\ (10, 1)\}\). Gaussian noise (zero mean, standard deviation 0.2) is added to the measurements to simulate realistic conditions.

\section{Results and Discussions}\label{Results and Discussions}
\subsection{The Decomposition of Nitrous Oxide}\label{ADoK-S Performance}
As outlined in Figure \ref{Fig:pi_adok_flowchart}, the first stage in deriving kinetic models with PI-ADoK is generating concentration profile models from dynamic experimental trajectories. To achieve this, we employ a genetic programming (GP) algorithm (using the implementation by Cranmer \cite{Cranmer_2023}) that constructs candidate expressions using the operator set \(\mathcal{P} = \{+,-,\div,\times,\exp\}\) and the variable set \(\mathcal{X} = \{t\}\), where \(t\) denotes time. This selection is motivated by our physical understanding of kinetic modeling and serves as an effective way to inject expert knowledge into the symbolic search.

\begin{figure}[htb!]
    \centering
    \includegraphics[width=0.9\textwidth]{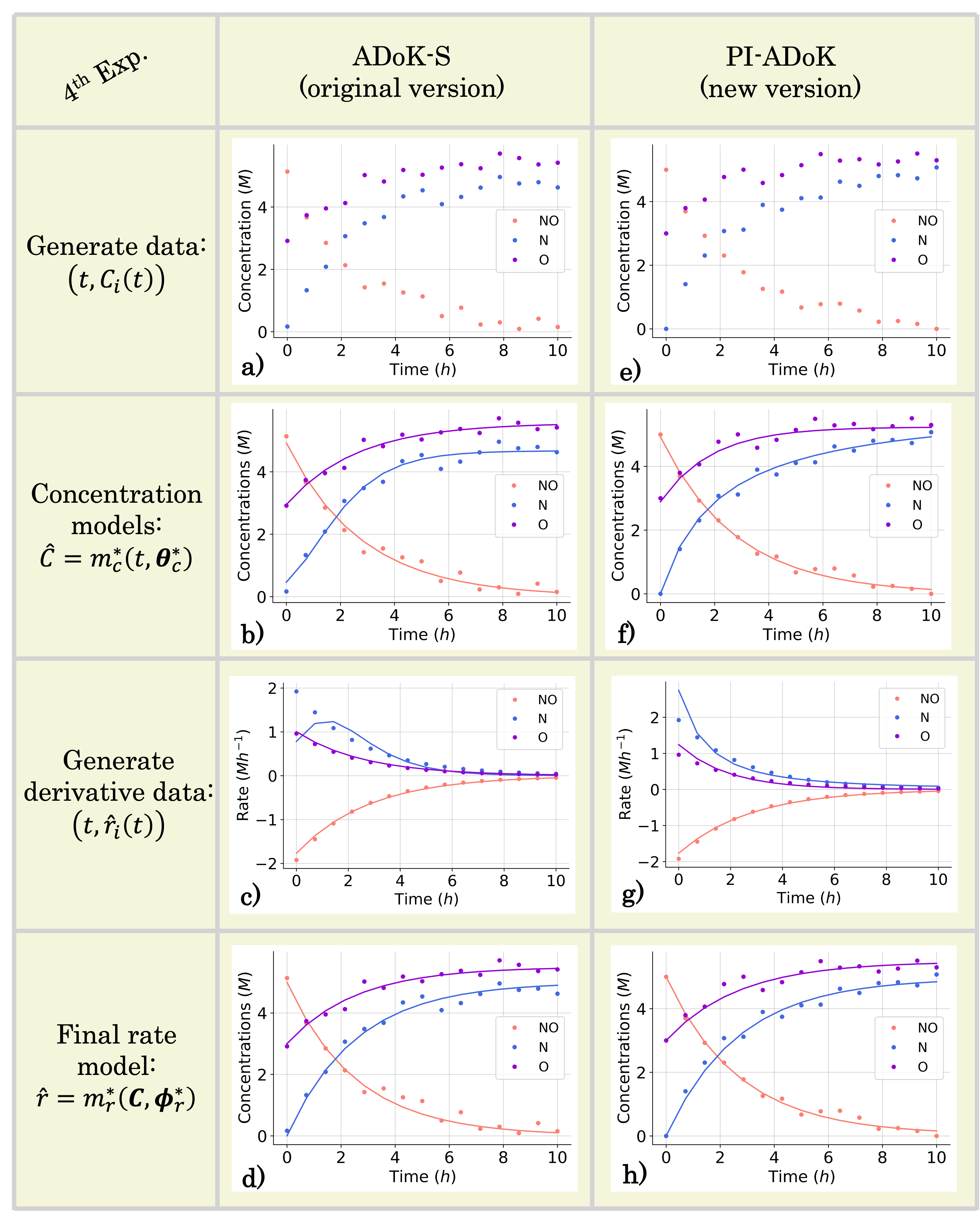}
    \caption{Illustration of the modeling workflow for the fourth experiment in the decomposition of nitrous oxide, comparing ADoK-S (left column) with PI-ADoK (right column). The first row \textbf{(a, e)} shows the in-silico concentration data. In the second row \textbf{(b, f)}, each method proposes concentration models that approximate these observations. The third row \textbf{(c, g)} displays the numerically differentiated rates inferred from the concentration models, and the final row \textbf{(d, h)} presents the final rate models. While both approaches capture the overall system dynamics, PI-ADoK enforces additional physical constraints (e.g., correct initial conditions and monotonic behavior), resulting in more accurate concentration profiles and improved rate estimates.}
    \label{Fig:pi_adok_results}
\end{figure}

In addition, we integrate a series of mathematical constraints derived from the in-silico data (see Figure \ref{Fig:pi_adok_results} (a) and (e)) to further guide the search. Specifically, our constraints ensure that: (i) the concentration models precisely reproduce the initial conditions, which are measured with high certainty; (ii) the models approach a chemical equilibrium over a sufficiently long time horizon (for instance, the concentrations should converge by 50 hours, so that the difference between \(t=50\) h and \(t=60\) h tends toward zero); (iii) the predicted concentrations remain non-negative, reflecting physical reality; and (iv) the reactant concentrations decrease monotonically while the product concentrations increase monotonically until equilibrium is reached.

It is important to note that although a closed-form solution to the underlying ODE system governing the reaction kinetics may not exist, the chosen construction rules have consistently demonstrated their capability to approximate both the concentration trajectories and the derived rate measurements effectively.

For this case study, we construct three concentration models for each experiment, specifically, \(\hat{C}_{NO,i}\), \(\hat{C}_{N,i}\), and \(\hat{C}_{O,i}\) for \(i \in \{1,2,\dots,5\}\), where \(NO\), \(N\), and \(O\) denote nitrous oxide, nitrogen, and oxygen, respectively. It is crucial to underscore that the development of each of these models is carried out autonomously. Although some might argue that this approach could yield models that violate essential physical principles such as mass conservation, our primary objective at this phase is to accurately approximate the system's rate measurements, even if a slight level of physical inconsistency is tolerated.

This section presents the results from the fourth experiment, which is representative of the overall methodology applied across all cases. Initially, the GP algorithm generates candidate concentration profile models for the species \textit{NO}, \textit{N}, and \textit{O} at various complexity levels (capped by the user). For example, the candidate concentration profiles for \textit{NO} in the fourth experiment are given by:

\begin{subequations}
\begin{gather}\label{Eq:2.3}
    \hat{C}_1(t) = p_1, \\
    \hat{C}_2(t) = \exp\left( p_1 - t \right), \\
    \hat{C}_3(t) = \frac{p_1}{p_2 + t}, \\
    \hat{C}_4(t) = \exp\left( p_1 - \frac{t}{p_2} \right), \\
    \hat{C}_5(t) = \frac{p_1-t}{p_2+t}, \\
    \hat{C}_6(t) = \frac{p_1-t}{p_2+t} + p_3.
\end{gather}
\end{subequations}

Here, each parameter \(p_i\) is estimated from the time-dependent concentration data for a given model, and \(\hat{C}_{i}(t)\) denotes the \(i^{\textrm{th}}\) proposed concentration model generated by PI-ADoK.

Following the construction of these concentration models, the next step involves parameter estimation aimed at minimizing the error between the model responses and the measured concentrations. Once the optimal parameters are determined, both the negative log-likelihood (NLL) and the Akaike information criterion (AIC) (see Eq. \eqref{eq:AIC}) are computed for each model. In this instance, model \(C_4(t)\) is selected to approximate the consumption rates for species \textit{NO} in the fourth experiment.

Figure \ref{Fig:pi_adok_results} displays the concentration profiles predicted by both PI-ADoK and ADoK-S. In panels (b) and (f), the concentration profiles from PI-ADoK and ADoK-S, respectively, are shown. Although both methods capture the overall dynamics, the models from ADoK-S exhibit noticeable discrepancies in the initial conditions, especially for nitrogen, whereas PI-ADoK, by enforcing the initial condition constraint, closely adheres to the true values.

Once the concentration profiles are validated, the corresponding rate estimates are derived through numerical differentiation. Panels (c) and (g) in Figure \ref{Fig:pi_adok_results} compare these estimated rates to the (hypothetical) rate measurements from the real system, \(\dot{x}(t)\). The inaccuracies in the initial conditions from ADoK-S result in rate estimates that significantly deviate from the expected values. In contrast, PI-ADoK yields rate estimates that are much more consistent with the system dynamics, underscoring the advantage of incorporating physical constraints.

In summary, the workflow demonstrated in this experiment begins with the generation of concentration profiles via GP, improved by constraints that enforce known physical behaviors (such as accurate initial conditions, attainment of equilibrium, non-negativity, and monotonic trends). These constraints lead to improved rate estimates through numerical differentiation. The comparative analysis clearly shows that PI-ADoK, by effectively incorporating these constraints, produces more reliable concentration models, as evidenced by the closer alignment of its rate estimates with the expected behavior. This advantage is critical for the accurate discovery of kinetic models in practical applications.

In alignment with the workflow depicted in Figure \ref{Fig:pi_adok_flowchart}, the next stage of PI-ADoK involves generating rate models using the same GP algorithm that was used to derive the concentration profiles. This stage unfolds iteratively, with the GP algorithm proposing candidate rate models that are refined to satisfy Eq. \eqref{Eq:1.1}. For this purpose, the expression construction rules are defined as \(\mathcal{P} = \{+,-,\div,\times\}\) and \(\mathcal{X} = \{C_{NO}, C_{N}, C_{O}\}\). These selections are based on our prior understanding of kinetic models and serve to inject expert knowledge into the symbolic search. Although the reaction rate is influenced solely by the concentrations of the species being measured, given that the experiments are conducted under constant temperature and volume, it is important to include \(C_N\) and \(C_O\) in the set \(\mathcal{X}\) since their potential influence cannot be ruled out a priori. Moreover, our experience allows us to narrow the operator set further, excluding, for example, trigonometric functions which are unlikely to appear in the rate expressions.

Based on the in-silico data, we also derive behavioral predictions for the rate models, which we encode as constraints in the GP algorithm. For concentration models, we enforce accurate prediction of the initial conditions; however, for rate models, we are as confident of our estimates at the beginning of the reaction as we are of our estimates at the end of the reaction. Analysis of the in-silico data reveals that the reactants’ concentrations decrease monotonically while the products’ concentrations increase monotonically. Therefore, we infer that the rate of consumption of reactants should remain always negative and monotonically increasing, whereas the rate of generation of products should be positive and monotonically decreasing.

Based on these construction rules and constraints, the GP algorithm proposes nine candidate rate model structures; for brevity, we present a select few:

\begin{subequations}
\begin{gather}\label{eq:rates}
    \hat{r}_1 = -k_1, \\
    \hat{r}_2 = -k_1C_{NO}, \\
    \hat{r}_3 = -k_1C_{NO} + k_2 + C_{NO}, \\
    \hat{r}_4 = -k_1\left(\left(C_{NO} - k_2\right) + \left(\frac{k_3}{k_4 + C_{NO}}\right)\right), \\
    \hat{r}_5 = -k_1\left(C_{NO} + \left(\frac{k_2}{k_3 + C_{NO}}\right)\right) - k_4, \\
    \hat{r}_6 = -k_1\left(C_{NO} + \left(\frac{k_2}{k_3 + C_{NO}}\right)\right) - \left(\frac{k_4}{k_5 - C_{NO}}\right).
\end{gather}
\end{subequations}

The parameters \(k_i\) for \(i \in \{1,2,\dots,5\}\) are estimated from the concentration data using dynamic parameter estimation. This estimation is achieved by solving Eq. \eqref{eq:PE Strong} with the ABC and LBFGS optimization algorithms. After computing the negative log-likelihood (NLL) and Akaike information criterion (AIC) for each candidate, the model with the lowest AIC is selected; in this case, \(\hat{r}_3\) is chosen, with its response illustrated in Figure \ref{Fig:pi_adok_flowchart}(d). For comparison, Figure \ref{Fig:pi_adok_flowchart}(h) shows the response of the selected model from ADoK-S after the initial five experiments (\(r = -k_1C_{NO}\)).

None of the candidate rate models in Eq. \eqref{eq:rates}, including \(\hat{r}_3\), match the data-generating rate model described in Eq. \eqref{eq:2.4}. Consequently, PI-ADoK must undergo an additional iteration using the Model-Based Design of Experiments (MBDoE) loop. In this loop, the top two models yielded by PI-ADoK, namely \(\hat{r}_3\) and \(\hat{r}_2\), are used to propose a discriminatory experiment by solving Eq. \eqref{Eq: MBDoE}.

The MBDoE procedure suggests running a sixth experiment with initial conditions \((C_{NO,0}, C_{N,0}, C_{O,0}) = (0.000, 1.522, 0.731)\) M. The new experiment follows the same sequence (generate, optimize, and select concentration models) to approximate the rates. Once the rates from the new experiment are computed, they are concatenated with the previous data, and the GP algorithm is re-run to generate, optimize, and select a refined set of rate models. The kinetic model selected by PI-ADoK after the sixth experiment, denoted as \(r^*\), is:

\begin{gather}\label{Eq:2.6}
    r^* = \frac{k_1C_{NO}^2}{1 + k_2C_{NO}}.
\end{gather}

Thus, after two iterations, PI-ADoK successfully uncovers a kinetic model (Eq. \eqref{Eq:2.6}) that is structurally identical to the data-generating model (Eq. \eqref{eq:2.4}). Notably, PI-ADoK required only six experiments to recover the model, whereas ADoK-S required 18 experiments: a reduction of 66.67\% in the experimental budget.

\begin{figure}[htb!]
    \centering
    \includegraphics[width=0.9\textwidth]{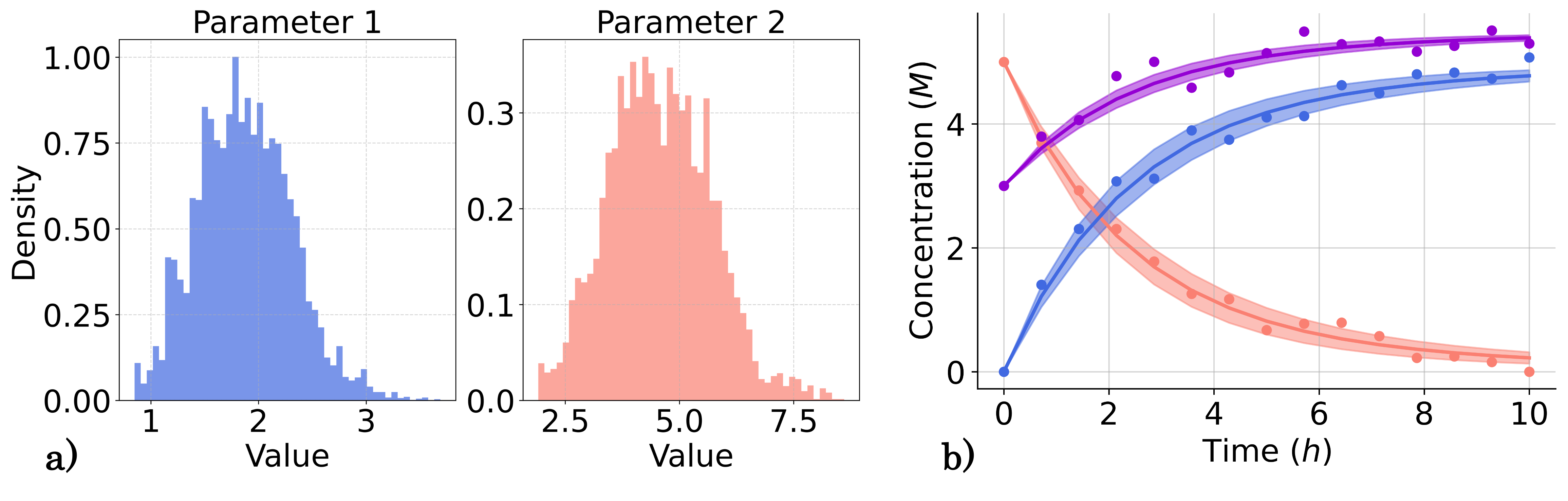}
    \caption{Illustration of the uncertainty quantification step for the selected kinetic model for the decomposition of nitrous oxide using PI-ADoK. \textbf{(a)} Posterior distributions for Parameter 1 ($k_1$) and Parameter 2 ($k_2$), estimated via Metropolis-Hastings sampling, indicating the range of plausible values after convergence. \textbf{(b)} The corresponding concentration predictions (solid lines) and associated uncertainty bands (shaded regions) are overlaid on the experimental data (dots). This visualization demonstrates how parameter uncertainty propagates through the model to influence the predicted concentration profiles.}
    \label{Fig:parameter_uncertainty_figure}
\end{figure}

Once the user is satisfied with the final model (or if the experimental budget is exhausted), the next step is to perform uncertainty quantification on the kinetic parameters. In our framework, this entails approximating the posterior distribution of these parameters, via a Metropolis-Hastings algorithm, and using the resulting samples to characterize the range of plausible parameter values. By propagating these posterior samples through the model’s governing equations, we can generate credible intervals for the predicted state trajectories, thereby gauging the reliability of model forecasts. Figure \ref{Fig:parameter_uncertainty_figure} (a) illustrates the posterior distributions of the parameters, where the mode is notably close to the data-generating values \((k_A = 2 \text{ M}^{-1} \text{h}^{-1}, k_B = 5 \text{ M}^{-1})\). Leveraging these posterior samples, we propagate parameter uncertainty through the kinetic model to estimate the corresponding uncertainty in the predicted concentration profiles. As shown in Figure \ref{Fig:parameter_uncertainty_figure} (b), we visualize the model’s predictions alongside the uncertainty bounds, extending up to three standard deviations. 

This final phase of uncertainty quantification is vital for informed decision-making in chemical process design and optimization. The distribution of potential outcomes offers insights into the robustness of model predictions, helping to identify whether further experiments are warranted to reduce uncertainty or whether alternative model forms should be considered. In essence, by combining PI-ADoK’s efficient model discovery with a rigorous uncertainty analysis, practitioners gain both a high-confidence kinetic model and a clear understanding of its predictive limitations.

\clearpage

\subsection{The Hydrodealkylation of Toluene}\label{Hydrodealkylation_results}
Starting from five initial experiments (as described in Section \ref{Hydrodealkylation}), PI-ADoK generates, optimizes, and selects concentration profile models for each species. To illustrate the process, we focus here on the first three experiments. Denoting \(\hat{C}_{i,j}\) as the model capturing the concentration dynamics of species \(i\) in experiment \(j\), we obtain:

\begin{subequations}
\begin{gather}\label{eq:conc_decom_strong}
    \hat{C}_{T,1}(t) = \frac{\exp(t)}{\exp(1.539t) - t}, \\
    \hat{C}_{H,1}(t) = 7.139 + \exp\bigl(1.590t - t\,\exp(t)\bigr), \\
    \hat{C}_{B,1}(t) = 3.015 - \exp\bigl(-0.686t\bigr), \\
    \hat{C}_{M,1}(t) = \frac{t - 0.027}{t + 0.939} + 3.054, \\[6pt]
    \hat{C}_{T,2}(t) = \frac{\exp(2.078)}{1.592 + t}, \\
    \hat{C}_{H,2}(t) = \exp\bigl(\exp(-0.284t) + 0.985\bigr) + 0.627, \\
    \hat{C}_{B,2}(t) = 4.373 - \frac{4.282}{\exp(0.475t)}, \\
    \hat{C}_{M,2}(t) = 4.944 - \exp\bigl(\exp(0.368) - 0.393t\bigr), \\[6pt]
    \hat{C}_{T,3}(t) = \exp\bigl(\exp(-0.252t - 0.242)\bigr) + 1.507, \\
    \hat{C}_{H,3}(t) = \exp\bigl(\exp(-0.229t)\bigr) - 0.081t, \\
    \hat{C}_{B,3}(t) = t \,\exp\bigl(\tfrac{t}{-2.576}\bigr) + 0.262t, \\
    \hat{C}_{M,3}(t) = \exp\Bigl(\exp\bigl(\tfrac{t}{t + 0.594}\bigr) - 1.472\Bigr).
\end{gather}
\end{subequations}

\begin{figure}[htb!]
    \centering
    \includegraphics[width=0.9\textwidth]{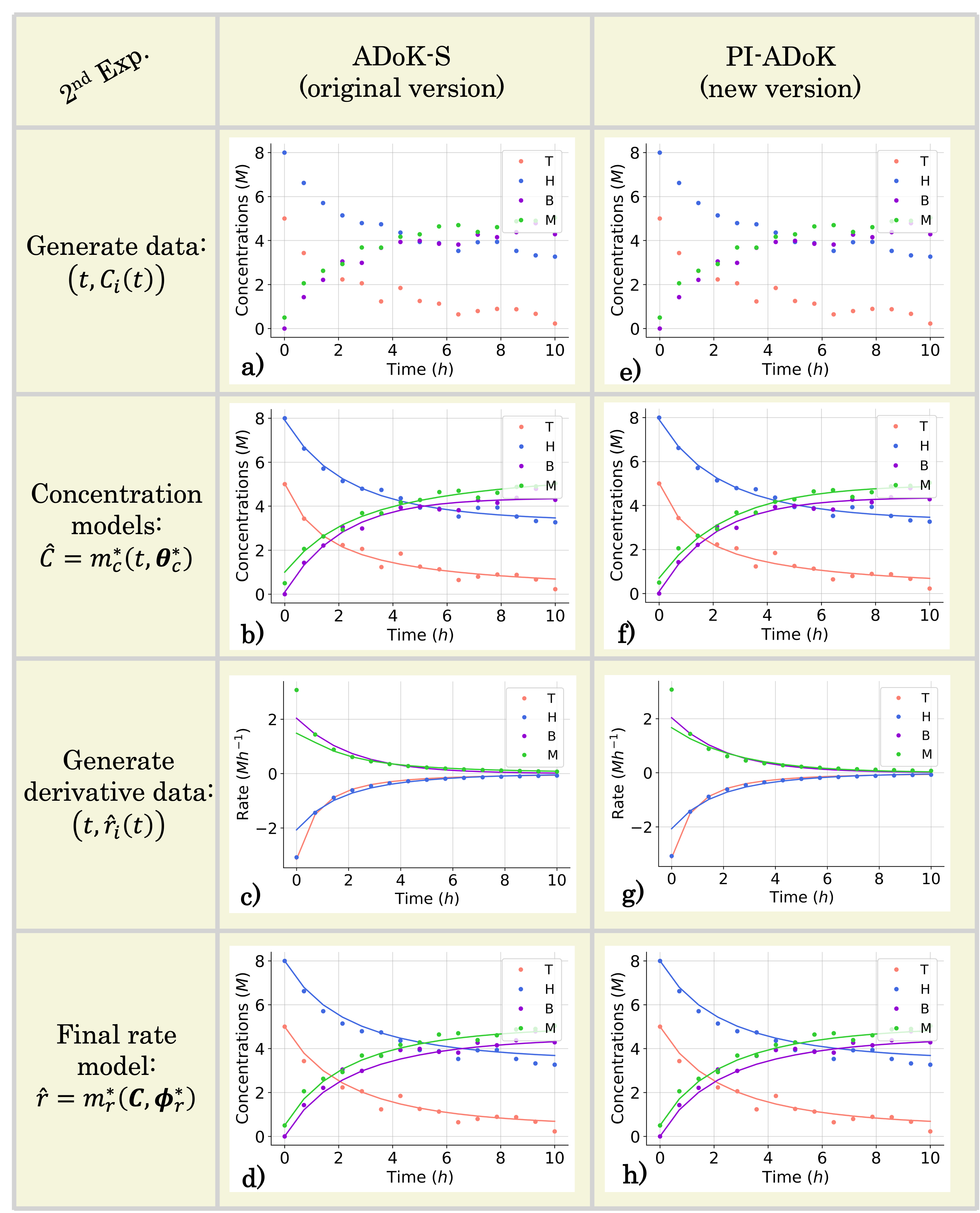}
    \caption{Illustration of the modeling workflow for the second experiment in the hydrodealkylation of toluene, comparing ADoK-S (left column) with PI-ADoK (right column). The first row \textbf{(a, e)} shows the in-silico concentration data. In the second row \textbf{(b, f)}, each method proposes concentration models that approximate these observations. The third row \textbf{(c, g)} displays the numerically differentiated rates inferred from the concentration models, and the final row \textbf{(d, h)} presents the final rate models. While both approaches capture the overall system dynamics, PI-ADoK enforces additional physical constraints (e.g., correct initial conditions and monotonic behavior), resulting in more accurate concentration profiles and improved rate estimates.}
    \label{Fig:pi_adok_results_2}
\end{figure}

Figure \ref{Fig:pi_adok_results_2} panels (a), (b), (e) and (f) display the synthetic measurements and the concentration surrogates chosen for the second experiment. Comparing panel (b) to panel (f), and focusing on the methane profile, reveals that the profile selected by PI-ADoK tracks the early-time dynamics better than the profile chosen by the benchmark method. This, once again, shows that the enforcement of constraints, particularly the initial condition-constraint, yields noticeably better results.

We next convert the fitted concentration profiles into pseudo-rate data by numerical differentiation. Panels (c) and (g) of Figure \ref{Fig:pi_adok_results_2} compare these numerical rates with the “true" (simulated) rates. Because the ADoK-S concentration surrogate already deviates slightly at \(t=0\) h, its derivative inherits those early-time errors. The PI-ADoK surrogate, by contrast, starts closer to the correct initial condition, so its differentiated curve follows the true early-time kinetics more closely. These improved rate estimates feed directly into the subsequent symbolic regression stage. When the resulting candidate rate laws are ranked by AIC, the two best models are:

\begin{subequations}
\begin{gather}\label{eq:rate_decom_strong}
    \hat{r}_1(t) = 0.049\,C_T\,C_H - 0.049\,C_B + 0.143, \\
    \hat{r}_2(t) = 0.049\,C_T\,C_H + 0.020\,C_T.
\end{gather}
\end{subequations}

Panels (d) and (h) of Figure \ref{Fig:pi_adok_results_2} show the performance of the selected models by ADoK-S and PI-ADoK, respectively. Despite the slight improvement of the rate estimations from PI-ADoK, we see that in this initial iteration, the performance of both models are almost identical.

Because neither of the models in Eq. \eqref{eq:rate_decom_strong} adequately captured the system dynamics, a MBDoE step was performed to propose a new experiment with initial conditions \(\bigl(C_T(0), C_H(0), C_B(0), C_M(0)\bigr) = \bigl(5.000, 6.954, 2.000, 2.660\bigr)\) M. Applying PI-ADoK to this sixth experiment yields the following concentration profiles:

\begin{subequations}
\begin{gather}\label{eq:conc_decom_strong_2}
    \hat{C}_{T,6}(t) = \exp\!\bigl(\exp(-0.139t + 0.524)\bigr) - 0.464, \\
    \hat{C}_{H,6}(t) = 2.170\,\exp\!\bigl(\exp(-0.138t + 0.151)\bigr) - 0.464, \\
    \hat{C}_{B,6}(t) = -0.050\,t^2 + 0.855\,t + 2.184, \\
    \hat{C}_{M,6}(t) = -0.051\,t^2 + 0.883\,t + 2.760.
\end{gather}
\end{subequations}

By numerically differentiating these concentration profiles to approximate the rate measurements for the sixth experiment, and concatenating the data with the previous experiments, PI-ADoK uncovers the following new rate models:

\begin{subequations}
\begin{gather}\label{eq:rate_decom_strong_3}
    \hat{r}_1(t) = \frac{0.272\,C_T^2\,C_H - 0.272\,C_T\,C_H\,C_B + 0.272}{\bigl(C_T + C_B\bigr)\bigl(C_T - C_B + 0.996\bigr) + 0.027}, \\
    \hat{r}_2(t) = \frac{C_T\,C_H}{3.610\,C_T + C_H\,C_B}.
\end{gather}
\end{subequations}

Because these newly proposed models still did not fully align with expectations, another MBDoE iteration suggested a seventh experiment with initial conditions \(\bigl(C_T(0), C_H(0), C_B(0), C_M(0)\bigr) = \bigl(5.000, 8.000, 0.696, 3.000\bigr)\) M. Reapplying PI-ADoK to this seventh experiment results in the concentration profiles:

\begin{subequations}
\begin{gather}\label{eq:conc_decom_strong_4}
    \hat{C}_{T,7}(t) = \frac{\exp(2.315)}{2.061 + t}, \\
    \hat{C}_{H,7}(t) = \exp\bigl(\exp(-0.238t + 0.501)\bigr) + 2.633, \\
    \hat{C}_{B,7}(t) = 5.063 - \exp\!\bigl(1.397 - \tfrac{t}{\exp(0.980)}\bigr), \\
    \hat{C}_{M,7}(t) = 7.035 - \frac{7.035 - t}{t + 1.718}.
\end{gather}
\end{subequations}

Finally, upon incorporating the rate measurements inferred from the seventh experiment, PI-ADoK converges on a rate model whose structure and parameter values closely match the data-generating rate equation:

\begin{equation}
    \hat{r}^* = \frac{2.256C_TC_H}{1 + 9.052C_B + 6.205C_T}.
\end{equation}

These results clearly illustrate the advantage of incorporating physical constraints into the model discovery process. Specifically, while PI-ADoK was able to recover a kinetic model that is structurally identical to the data-generating model after only 7 experiments, ADoK-S required 16 experiments to achieve the same outcome. This represents a reduction of 56.25\% in the number of experiments needed, underscoring the efficiency gains from integrating constraints. By narrowing the search space and steering the GP algorithm toward physically plausible solutions, the added constraints not only enhance model accuracy but also significantly lower the experimental burden, a crucial benefit in resource-limited experimental settings.

Once the rate law is accepted, or further experiments are no longer feasible, we quantify parameter uncertainty. A Metropolis–Hastings algorithm samples the posterior distribution of the kinetic parameters, outlining the full range of plausible values. Propagating these samples through the model produces credible intervals for the concentration trajectories and hence a direct measure of prediction reliability. Figure \ref{Fig:parameter_uncertainty_figure_3} (a) shows the posterior densities; their modes lie close to the true parameters \(k_A = 2\,\mathrm{M}^{-1}\mathrm{h}^{-1}\), \(K_B = 9\,\mathrm{M}^{-1}\), and \(K_C = 5\,\mathrm{M}^{-1}\). Panel (b) overlays the predicted concentrations with \(\pm 3\sigma\) uncertainty intervals obtained from the same posterior distributions.

\clearpage

\begin{figure}[htb!]
    \centering
    \includegraphics[width=0.9\textwidth]{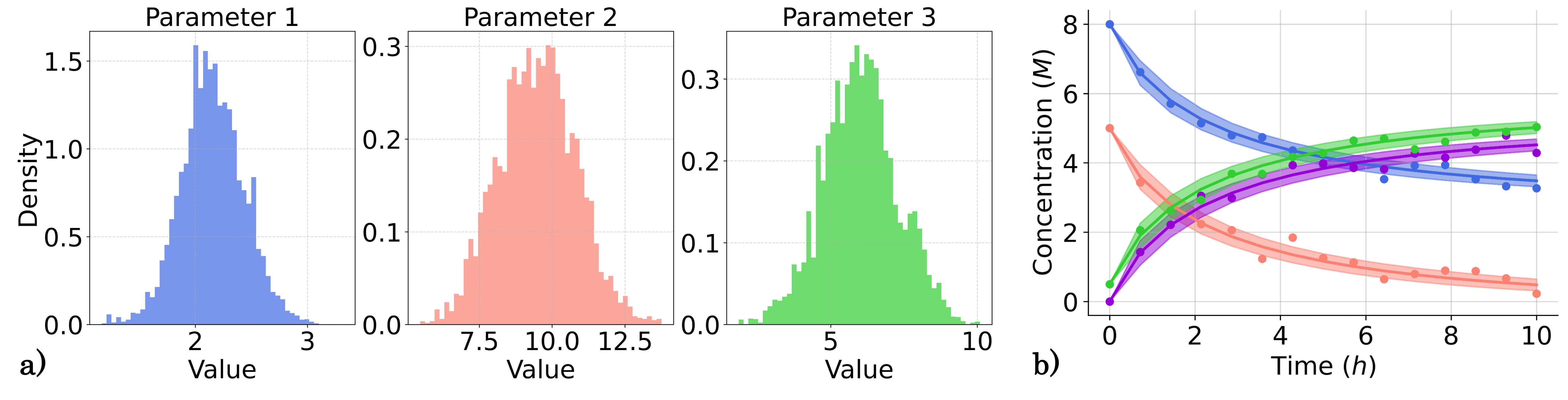}
    \caption{Illustration of the uncertainty quantification step for the selected kinetic model for the hydrodealkylation of toluene using PI-ADoK. \textbf{(a)} Posterior distributions for Parameter 1 ($k_1$), Parameter 2 ($k_2$) and Parameter 3 ($k_3$), estimated via Metropolis-Hastings sampling, indicating the range of plausible values after convergence. \textbf{(b)} The corresponding concentration predictions (solid lines) and associated uncertainty bands (shaded regions) are overlaid on the experimental data (dots). This visualization demonstrates how parameter uncertainty propagates through the model to influence the predicted concentration profiles.}
    \label{Fig:parameter_uncertainty_figure_2}
\end{figure}

\subsection{The Theoretical Isomerization Reaction}
In this subsection, we present the results for the isomerization case study using PI-ADoK. Based on five initial experiments (described in Section \ref{Isomerization}), PI-ADoK generated, optimized, and selected candidate concentration profile models for each species across the experiments. Here, \(\hat{C}_{i,j}(t)\) denotes the model that characterizes the dynamic evolution of the concentration of species \(i\) in experiment \(j\):

\begin{subequations}
\begin{align}
\hat{C}_{A,1}(t) = 0.558 + \exp\bigl(0.602 - t\bigr), \\
\hat{C}_{B,1}(t) = 1.450 - \frac{1.507}{\exp(t)}, \\
\hat{C}_{A,2}(t) = \exp\Bigl(\exp\bigl(-0.067t + 0.836\bigr)\Bigr), \\
\hat{C}_{B,2}(t) = \frac{t}{\exp(-0.388)\,\exp(0.080t)}, \\
\hat{C}_{A,3}(t) = \exp\Bigl(\frac{t}{0.402} - \exp(t)\Bigr) + 1.304, \\
\hat{C}_{B,3}(t) = 2.803 - \exp\Bigl(-0.132\,\exp(t)\Bigr), \\
\hat{C}_{A,4}(t) = 0.057t^2 - 1.123t + 9.837, \\
\hat{C}_{B,4}(t) = -0.057t^2 + 1.123t + 2.060, \\
\hat{C}_{A,5}(t) = \exp\Bigl(\exp\Bigl(\frac{\exp(-0.098t)}{1.169}\Bigr)\Bigr), \\
\hat{C}_{B,5}(t) = -0.065t^2 + 1.244t + 1.084.
\end{align}
\end{subequations}

\begin{figure}[htb!]
    \centering
    \includegraphics[width=0.9\textwidth]{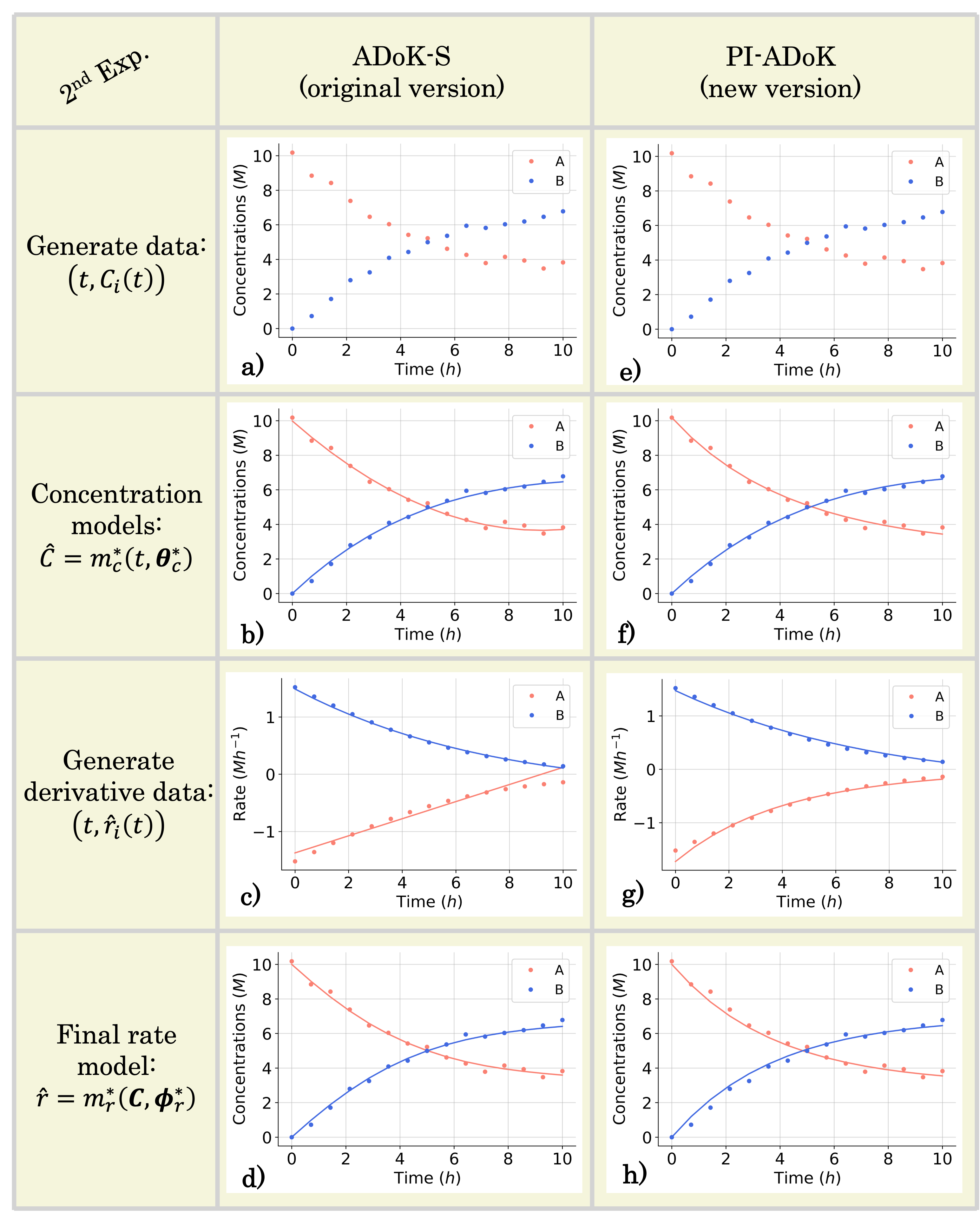}
    \caption{Illustration of the modeling workflow for the second experiment in the hypothetical isomerization study, comparing ADoK-S (left column) with PI-ADoK (right column). The first row \textbf{(a, e)} shows the in-silico concentration data. In the second row \textbf{(b, f)}, each method proposes concentration models that approximate these observations. The third row \textbf{(c, g)} displays the numerically differentiated rates inferred from the concentration models, and the final row \textbf{(d, h)} presents the final rate models. While both approaches capture the overall system dynamics, PI-ADoK enforces additional physical constraints (e.g., correct initial conditions and monotonic behavior), resulting in more accurate concentration profiles and improved rate estimates.}
    \label{Fig:pi_adok_results_3}
\end{figure}

Figure \ref{Fig:pi_adok_results_3} panels (a), (b), (e), and (f) show the in-silico data and the concentration surrogates selected in the second experiment. A direct comparison of panels (b) and (f) for species B shows that PI-ADoK reproduces the approach to equilibrium more accurately than ADoK-S: evidence that the equilibrium enforcement constraint improves the fit.

Differentiating the surrogates yields pseudo-rate data. Panels (c) and (g) plot these numerical rates against the true (simulated) rates of generation and consumption of the products and reactants, respectively. Because the ADoK-S surrogate drifts between \(t=8\) h and \(t=10\) h, that error is magnified in the derivative space; the PI-ADoK surrogate, which approaches equilibrium smoothly, produces rates that adhere closely to the ground truth. With these sharper estimates PI-ADoK subsequently recovers a rate law that almost matches the data-generating kinetics:

\begin{equation}
    \hat{r}^* = \frac{7.689C_A-1.896C_B}{4.053C_A+1.608C_B+5.943}.
\end{equation}

An important observation is that PI-ADoK dramatically reduces the experimental burden required to recover the true kinetic model. In our study, while the unconstrained ADoK-S approach necessitated 16 experiments to converge on the data-generating model, PI-ADoK achieved this with only the 5 initial experiments: a reduction of 68.75\% in the number of experiments. This substantial decrease highlights, just like in the other case studies, the efficacy gain of incorporating physical constraints into the discovery process, as these constraints effectively direct the search toward regions of the model space that are both accurate and physically plausible.

\begin{figure}[htb!]
    \centering
    \includegraphics[width=0.9\textwidth]{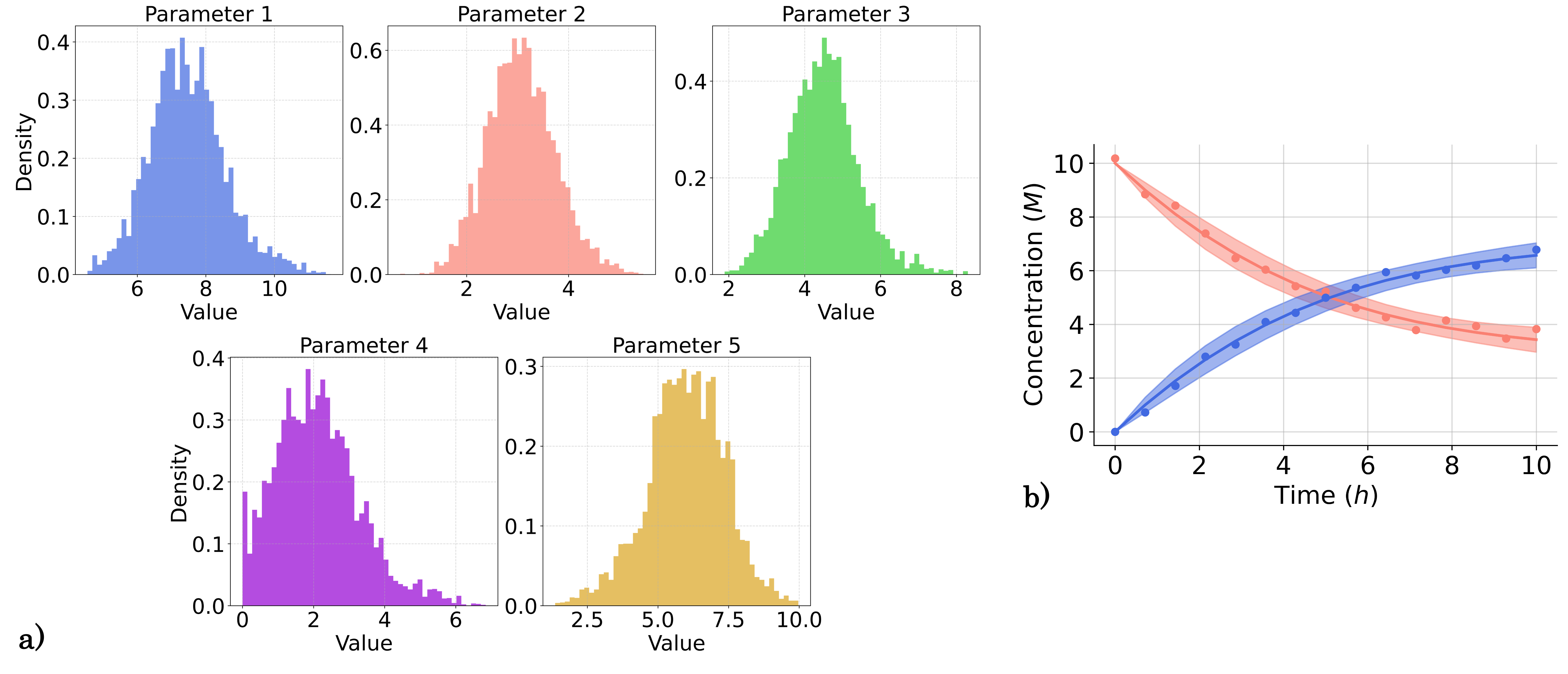}
    \caption{Illustration of the uncertainty quantification step for the selected kinetic model for the hypothetical isomerization study using PI-ADoK. \textbf{(a)} Posterior distributions for Parameter 1 ($k_1$), Parameter 2 ($k_2$) Parameter 3 ($k_3$), Parameter 4 ($k_4$) and Parameter 5 ($k_5$), estimated via Metropolis-Hastings sampling, indicating the range of plausible values after convergence. \textbf{(b)} The corresponding concentration predictions (solid lines) and associated uncertainty bands (shaded regions) are overlaid on the experimental data (dots). This visualization demonstrates how parameter uncertainty propagates through the model to influence the predicted concentration profiles.}
    \label{Fig:parameter_uncertainty_figure_3}
\end{figure}

After the final rate law has been accepted, or when no additional experimentation is possible, we assess parameter uncertainty.  Using a Metropolis–Hastings algorithm, we draw from the posterior distribution of the kinetic coefficients, thereby mapping the full spectrum of plausible values. Running these samples through the model yields credible intervals for the concentration profiles, providing a quantitative gauge of prediction reliability.  Figure \ref{Fig:parameter_uncertainty_figure_3} (a) displays the posterior densities, whose modes align closely with the true parameters \(k_A = 7\,\mathrm{M\,h^{-2}},\; k_B = 3\,\mathrm{M\,h^{-2}},\; k_C = 4\,\mathrm{h^{-1}},\; k_D = 2\,\mathrm{h^{-1}},\; k_E = 6\,\mathrm{M\,h^{-1}}\). Panel (b) shows the predicted concentration profiles with the \(\pm 3\sigma\) confidence intervals derived from these posterior samples.

\clearpage

\section{Conclusions}\label{Conclusions}
In this work, we introduced the Physics-Informed Automated Discovery of Kinetics (PI-ADoK) framework, an enhanced data-driven approach for discovering kinetic rate models from noisy concentration measurements. By integrating physical constraints directly into the genetic programming (GP) algorithm, PI-ADoK guides the model discovery process toward solutions that are not only statistically optimal but also physically plausible. Unlike traditional mechanistic models that require extensive prior knowledge and resource-intensive development, or black-box methods that sacrifice interpretability, our approach offers a transparent, efficient, and interpretable pathway to kinetic model identification.

A key innovation in PI-ADoK is the incorporation of constraints based on fundamental chemical principles -- such as ensuring accurate initial conditions, enforcing equilibrium behavior, maintaining non-negativity, and preserving monotonic trends. These constraints narrow the search space and focus computational effort on the most promising regions, which, as our case studies demonstrate, leads to significant reductions in experimental effort. For example, while the unconstrained ADoK-S framework required up to 16 experiments to converge on the data-generating kinetic model in one case study, PI-ADoK was able to recover an equivalent model with only 5 experiments -- a reduction of 68.75\% in experimental requirements. This dramatic improvement underscores the power of embedding physical insights into the discovery task.

Our comparative evaluations, conducted on several catalytic reaction systems -- including the decomposition of nitrous oxide, the hydrodealkylation of toluene, and a theoretical isomerization reaction -- demonstrate that the integration of physical constraints not only improves the accuracy of concentration and rate estimates but also enhances the overall reliability of the kinetic models. The experimental results, summarized in Table~\ref{Table:SummarizedResults}, highlight that PI-ADoK consistently recovers kinetic models that closely mirror the true dynamics of the systems under investigation, while also reducing the experimental burden.

\begin{table}[htb!]
\caption{The summarized results of the performance of PI-ADoK and ADoK-S against all three case studies explored.}
\begin{tabular}{p{0.22\textwidth}p{0.22\textwidth}p{0.22\textwidth}p{0.22\textwidth}}
\hline
    & Hypothetical isomerization reaction &  Decomposition of nitrous oxide & Hydrodealkylation of toluene \\
\hline
    Number of experiments -- PI-ADoK & 5 & 6 & 7\\

    Number of experiments -- ADoK-S & 16 & 18 & 16\\

    Data efficiency gain & 68.75\%  & 66.67\%  & 56.25\% \\

    Data-generating kinetic model & $\frac{7C_A-3C_B}{4C_A+2C_B+6}$ & $\frac{2C_{N_2O}^2}{1+5C_{N_2O}}$ & $\frac{2C_TC_H}{1+9C_B+5C_T}$ \\

    Rate model uncovered -- PI-ADoK & $\frac{7.689C_A-1.896C_B}{4.053C_A+1.608C_B+5.943}$ & $\frac{1.842C_{N_2O}^2}{1+4.598C_{N_2O}}$ & $\frac{2.256C_TC_H}{1 + 9.052C_B + 6.205C_T}$ \\

    Rate model uncovered -- ADoK-S & $\frac{8.365C_A-2.002C_B}{4.546C_A+1.634C_B+6.596}$ & $\frac{2.286C_{N_2O}^2}{1+5.792C_{N_2O}}$ & $\frac{2.100C_TC_H}{1 + 9.350C_B + 5.342C_T}$\\
\hline
\end{tabular}
\label{Table:SummarizedResults}
    \centering
\end{table}

In addition to the improved efficiency and model fidelity, PI-ADoK lays the groundwork for a comprehensive uncertainty quantification process. Once a model is deemed satisfactory or when the experimental budget is exhausted, the framework facilitates uncertainty analysis by propagating the uncertainty in the kinetic parameters through the kinetic model. This allows for the estimation of uncertainty intervals for predicted concentrations, thus providing valuable insights into the reliability of model forecasts and aiding further decision-making.

While our results are promising, we recognize that the success of any data-driven approach not only depends on the quality of the experimental data but also on the effective tuning of the hyperparameters that govern the imposed physical constraints. In our current implementation, these hyperparameters have been set statically; however, future work could explore dynamic hyperparameter tuning strategies. For example, one could begin with more relaxed constraints to promote model diversity during the early iterations, and then gradually enforce stricter constraints as the search converges toward promising regions. Such adaptive tuning could further enhance model robustness and reduce the experimental burden.

Moreover, it would be valuable to systematically evaluate alternative sampling techniques -- benchmarking methods such as Hamiltonian Monte Carlo against Metropolis-Hastings -- to assess their relative efficiency and accuracy in propagating uncertainty. Additionally, a deeper investigation into the relative importance of different constraints could yield insights into which physical principles are most critical for guiding the discovery task. This understanding would enable a more targeted integration of expert knowledge, ultimately leading to improved model fidelity and broader applicability of the framework across diverse systems.

In summary, by combining automated symbolic regression with physics-based constraints and robust uncertainty quantification, PI-ADoK represents a significant improvement in the development of reliable, data-efficient kinetic models. This work opens new avenues for the safe and efficient design of chemical processes, and we anticipate that future enhancements -- such as dynamic hyperparameter tuning and further integration of domain-specific knowledge -- will continue to improve its performance and applicability.

\section*{Author Contributions}

\textbf{Miguel Ángel de Carvalho Servia:} Conceptualization, formal analysis, investigation, methodology, project administration, software development, validation, visualization, writing (original draft), and writing (review and editing).

\textbf{Ilya Orson Sandoval:} Methodology, software development, and writing (review and editing).

\textbf{King Kuok (Mimi) Hii:} Conceptualization, formal analysis, funding acquisition, supervision, writing (original draft), and writing (review and editing).

\textbf{Klaus Hellgardt:} Conceptualization, formal analysis, funding acquisition, supervision, and writing (review and editing).

\textbf{Dongda Zhang:} Conceptualization, funding acquisition and supervision.

\textbf{Ehecatl Antonio del Rio Chanona:} Conceptualization, formal analysis, funding acquisition, methodology, project administration, supervision, and writing (review and editing).

\section*{Declaration of Competing Interest}
The authors declare that they have no known competing financial interests or personal relationships that could have appeared to influence the work reported in this paper.

\section*{Acknowledgments and Funding}
This work was supported by the EPSRC Centre of Doctoral Training for Next Generation Synthesis \& Reaction Technology (rEaCt) funding grant EP/S023232/1.

\section*{Appendix A. Supplementary Information}


The code used to produce all results and graphs shown in this work is available upon request.




\bibliographystyle{unsrtnat}
\bibliography{references.bib}

\end{document}